\documentclass[letterpaper, 10 pt, conference]{ieeeconf}  

\usepackage{mathrsfs}
\usepackage{amssymb}
\usepackage{amsmath,bm}
\usepackage{bbold}

\usepackage{mathrsfs}
\usepackage{graphicx}
\usepackage{enumerate}
\usepackage{eurosym}
\usepackage{amssymb}
\usepackage{amsmath,bbm}
\usepackage{amsfonts}
\usepackage{epstopdf}
\usepackage{epsf}
\usepackage{subfig}
\usepackage{psfrag}
\usepackage{graphics}
\usepackage{color} 
\usepackage{cite}
\usepackage{array,multirow,pbox}
\usepackage{tabularx,makecell,hhline}
\usepackage{soul}

\newtheorem{remark}{Remark}

\graphicspath{{../figures/}}

\IEEEoverridecommandlockouts                              

\title{\LARGE \bf
Optimization-Based Resiliency Verification in Microgrids via Maximal Adversarial Set Characterization
}

\author{Nawaf Nazir$^*$, Thiagarajan Ramachandran, Saptarshi Bhattacharya, Ankit Singhal, \\Soumya Kundu, and Veronica Adetola
\thanks{The authors are with the Electricity Infrastructure and Buildings Division at the Pacific Northwest National Laboratory, Richland, WA 99354, USA. S. Kundu holds a joint appointment as an Adjunct Faculty within the Department of Electrical and Biomedical Engineering at the University of Vermont, Burlington, VT 05405, USA.}%
\thanks{*Corresponding Author. Email:  {\tt\small nawaf.nazir@pnnl.gov}.}
}

\begin{document}

\maketitle
\thispagestyle{empty}
\pagestyle{empty}

\begin{abstract}

Critical energy infrastructures are increasingly relying on advanced sensing and control technologies for efficient and optimal utilization of flexible energy resources. Algorithmic procedures are needed to ensure that such systems are designed to be resilient to a wide range of cyber-physical adversarial events. 
This paper provides a robust optimization framework to quantify the range of adversarial perturbations that a system can accommodate without violating pre-specified resiliency metrics. An inner-approximation of the set of adversarial events which can be mitigated by the availabe flexibility is constructed using an optimization based approach. The proposed algorithm is illustrated on an islanded microgrid example: a modified IEEE 123-node feeder with distributed energy resources. Simulations are carried out to validate that the resiliency metrics are met for any event sampled from the constructed adversarial set for varying levels of available flexibility (energy reserves).

\end{abstract}

\section{INTRODUCTION}
\textit{Decarbonization}, \textit{digitalization}, and \textit{decentralization} are driving the evolution of the energy infrastructures worldwide in the 21st century \cite{di2018decarbonization}. There are incentives towards low carbon solutions (including, but not limited to, renewable energy resources), most of which are located at the customer sites (e.g., rooftop solar panels). On the other hand, advanced sensing and control technologies, coupled with ubiquitous communication infrastructure, are creating a digital platform for efficient and optimal utilization of distributed energy resources. Fueled by these and other drivers, microgrids are fast emerging as viable energy infrastructure. Microgrids are a group of flexible energy resources operating together locally as a single controllable entity, satisfying certain reliability, power quality, and adequacy standards \cite{farrokhabadi2019microgrid}. 
%
Safe and reliable operation of microgrids via optimal coordination of the flexible energy resources has drawn much attention in recent works \cite{parhizi2015state}. 
%
%
Several works, including~\cite{anderson2017increasing}, have shown how microgrids with multiple energy sources can minimize the cost of energy and increase the resilience of energy supply to critical loads. %
In~\cite{hussain2019impact}, data-driven methods have been used to study the impact of uncertainties in the renewable generation on microgrid resiliency, whereas in~\cite{wu2019microgrid}, resiliency metrics are incorporated into the microgrid planning problem. A robust optimization framework for microgrid operation is presented in~\cite{liu2017robust}, which calculates the reserves required to ensure robustness under modeled uncertainties. 

The resiliency of energy systems, e.g., microgrids, against various cyber and physical adversarial events, including malicious attacks, have been studied in the literature \cite{dehghani2021cyber,wang2019deep,dehghani2021blockchain,sahoo2018stealth}. One type of such attacks includes injecting measurement errors to renewable generation output. Such errors can be constant offsets or potentially time-varying, and left untreated, may eventually lead to ramp-induced attacks \cite{che2019impact}. Another class of attacks include \textit{replay attacks} \cite{habibi2020detection}, whereby the measured renewable generation output is maliciously changed to reflect the true generation at a previous time stamp, thus impacting dispatch decisions. A third class of attacks entails corruption of state estimation, whereby the estimated physical quantities required for dispatch decisions (such as load setpoints) are maliciously changed \cite{kushal2018risk,banerjee2021online}. Specifically, load-setpoint change attacks often cause maximal impact when affected during specific hours of the day (for example, during 12\,-\,7\,PM, when the load typically rises to hit an evening peak). A final class of attacks is coordinated attacks, whereby an attacker often has access to multiple critical information of the system, and can maliciously pose threats in a coordinated manner, from multiple sources (such as malicious under-estimation of load coupled with over-estimation of renewable capacity). Examples of such cyber-physical attacks can be found in~\cite{sun2018cyber,he2021tri,che2019impact}. 
In addition to attacks, other physical events such as generator outage, loss of solar generation due to cloud coverage, etc., could also lead to violations of critical operational constraints (such as serviceability of critical loads), as considered in this work.

The main contribution of the paper is to identify the set of cyber-physical adversarial events that the microgrid is guaranteed to be resilient against, via optimal use of available flexible energy resources. An optimization problem is set up to construct an inner approximation of the tolerable adversarial set.
The rest of the paper is organized as follows. Sec.\,\ref{sec:baseline} presents the microgrid model that will be used in this paper and presents baseline optimization results without considering the uncertainty due to adversarial events. Sec.\,\ref{sec:verification} provides the robust version of the microgrid optimization problem that calculates optimal reserves required to manage the adversarial events. Sec.\,\ref{sec:adv_set} provides the adversarial set characterization technique that is used to determine an inner approximation of the set of adversarial events that the microgrid can handle. Simulation results showing various case studies on the baseline model, robust model and adversarial set characterization are shown in Sec.\,\ref{sec:results}. Finally, Sec.\,\ref{sec:concl} provides the conclusions and the scope for future work.

\section{BASELINE OPTIMIZATION}
\label{sec:baseline}
Let us begin by presenting a formulation of the optimal dispatch problem in a microgrid with various distributed energy resources, including solar photovoltaics (PVs), diesel generators (DGs), and energy storage (ES) units. We introduce a few notations for the following mathematical description. Let $\mathcal{N}$ denote the set of all bus indices, and $\mathcal{L}\subseteq\mathcal{N}\times\mathcal{N}$ denote the set of all branches. We use $V_{n,k}$ to denote the multi-phase voltage vector at bus $n\in\mathcal{N}$, at time $k$. The net apparent power injection at bus $n$, at time $k$, is denoted by $S_{n,k}^{\text{net}}=P_{n,k}^{\text{net}}+jQ_{n,k}^{\text{net}}$, where $P_{n,k}^{\text{net}}$ and $Q_{n,k}^{\text{net}}$ are the respective active and reactive power components. Likewise, the Table\,\ref{tab:notations} lists the symbols denoting the apparent, active, and reactive power associated with each of the energy resources and loads at bus $n$, at time $k$\,.
\begin{table}[htpb]
    \centering
    \begin{tabular}{c|c|c}
    \hline
         \textsc{apparent, active, reactive} & \textsc{unit} & \textsc{convention}\\
         \hhline{---}
         $\lbrace S_{n,k}^\text{pv}, P_{n,k}^\text{pv}, Q_{n,k}^\text{pv}\rbrace$ & solar PV & injection\\
         $\lbrace S_{n,k}^\text{dg}, P_{n,k}^\text{dg}, Q_{n,k}^\text{dg}\rbrace$ & diesel generator & injection\\
         $\lbrace S_{n,k}^\text{es}, P_{n,k}^\text{es}, Q_{n,k}^\text{es}\rbrace$ & energy storage & injection\\
         $\lbrace S_{n,k}^\text{load},P_{n,k}^\text{load},Q_{n,k}^\text{load}\rbrace$ & load & consumption\\
         \hline
    \end{tabular}
    \caption{Notations used for nodal power exchange}
    \label{tab:notations}
\end{table}
Let $E_{n,k}^\text{es}$ be the storage state of charge (SoC). Furthermore, let $i_{n,m,k}$ be the complex vector denoting the multi-phase current flowing from bus $n$ to bus $m$ (i.e., $n\rightarrow m$), at time $k$. Let $S_{n,m,k},\,P_{n,m,k}$ and $Q_{n,m,k}$ denote, respectively, the apparent, active, and reactive branch power-flow from $n\rightarrow m$\,, at time $k$\,. Let $Z_{n,m}\!=\!Z_{m,n}$ be the impedance matrix of the branch $(n,m)\!\in\!\mathcal{L}$. Furthermore, we define (with $\left(\cdot\right)^*$ denoting conjugate transpose):
\begin{align*}
    W_{n,k}:=V_{n,k}\cdot V_{n,k}^*\,,\quad I_{n,m,k}:=i_{n,m,k}\cdot i_{n,m,k}^*\,.
\end{align*}


The distribution network power-flow equations can be expressed in the Branch Flow Model (BFM) \cite{gan2014convex} as:
\begin{subequations}
\label{eq:pf_bfm}
\begin{align}
\begin{bmatrix}
W_{n,k} & S_{n,m,k}\\
S_{n,m,k}^{*} & I_{n,m,k}\end{bmatrix}=
\begin{bmatrix}
V_{n,k}\\
i_{n,m,k}\end{bmatrix}
\begin{bmatrix}
V_{n,k}\\
i_{n,m,k}\end{bmatrix}^{*} &\label{eq:P1_BFM}\\
W_{n,k}=W_{m,k}-(S_{n,m,k}Z_{n,m}^*+Z_{n,m}S_{n,m,k})&\notag\\
+Z_{n,m}I_{n,m,k}Z_{n,m}^*&\label{eq:P1_volt_rel}\\
\!\!S^{\text{net}}_{n,k}+\text{diag}(S_{n,m,k}\!-\!Z_{n,m} I_{n,m,k})=\!\!\!\sum_{o:m\rightarrow o}\!\!\!\text{diag}( S_{m,o,k})& \label{eq:P1_power_balance}
\end{align}
\end{subequations}
The nonlinearity of the power-flow equations in \eqref{eq:pf_bfm} render the resulting optimal dispatch problem non-convex (and NP-hard). While convex relaxations of the power-flow have been proposed in the literature (see \cite{nazir2020optimal} and the references therein), in this paper, we adopt a linearized power-flow model which typically yields acceptable solutions \cite{bolognani2015fast}. In particular, the linearized BFM model \cite{gan2014convex} is given as below:
\begin{subequations}
\label{eq:pf_linear}
\begin{align}
    0&=W_{n,k}-W_{m,k}+(S_{n,m,k}Z_{n,m}^*+Z_{n,m}S_{n,m,k})\label{eq:P2_volt_rel}\\
0&=\text{diag}(S_{n,m,k})-\sum_{o:m\rightarrow o}\text{diag}(S_{m,o,k})+S^{\text{net}}_{n,k}\label{eq:P2_power_balance}
\end{align}
\end{subequations}
Next, based on the convention in Table\,\ref{tab:notations}, the net active and reactive power injections at each bus $n$ satisfy the relations:
\begin{subequations}
\begin{align}
    P^{\text{net}}_{n,k}&=P^\text{pv}_{n,k}+P^\text{dg}_{n,k}+P^\text{es}_{n,k}-P^\text{load}_{n,k} \label{eq:P1_node_real_balance}\\
Q^\text{net}_{n,k}&=Q^\text{pv}_{n,k}+Q^\text{dg}_{n,k}+Q^\text{es}_{n,k}-Q^\text{load}_{n,k}\label{eq:P1_node_reactive_balance}\
\end{align}
\end{subequations}

The power system operational reliability and safety dictate imposing certain constraints on the bus voltages and the line flows, as follows:
\begin{subequations}
\label{eq:pf_constraints}
\begin{align}
    &\left|\text{diag}(S_{n,m,k})\right|\leq \overline{S}_{n,m} \label{eq:P1_line_const}\\
&\underline{V}^2\leq \text{diag}(W_{n,k})\leq \overline{V}^2 \label{eq:P1_volt_const}
\end{align}
\end{subequations}
where $\underline{V}$ and $\overline{V}$ are, respectively, the maximum and minimum allowable voltage limits (e.g., specified by ANSI C84.1 Standard), while $\overline{S}_{n,m}$ is the maximum limit on the line-flow.

The inequality constraints that govern the solar PV and diesel power generation is given by the following:
\begin{align}
\left|S^\text{pv}_{n,k}\right|\leq \overline{S}^\text{pv}_n\,,\quad \left|S^\text{dg}_{n,k}\right|\leq \overline{S}^\text{dg}_n\,, \label{eq:gen_constraints}
\end{align}
where $\overline{S}^\text{pv}_n$ denotes the inverter capacity at bus $n$, and $\overline{S}^\text{dg}_n$ is the rated size of the diesel generator at bus $n$.

Finally, the energy storage units are associated with a state-of-charge (SoC) dynamics, as well as limits on the charging/discharging rates, limits on the SoC (i.e., storage size), and the inverter capacity, as follows:
\begin{subequations}\label{eq:es_constraints}
\begin{align}
E_{n,k+1}^\text{es}&=E_{n,k}^\text{es}-P^\text{es}_{n,k}\Delta t \label{eq:P1_battery_power_rel}\\
\quad \underline{E}_n^\text{es} &\leq E_{n,k}^\text{es}\leq \overline{E}_n^\text{es} \label{eq:P1_SOC_limit}\\
-\overline{P}_n^\text{es} &\leq P^\text{es}_{n,k}\leq \overline{P}_n^\text{es} \label{eq:P1_Pd_limit}\\
\left|S^\text{es}_{n,k}\right|&=\sqrt{(P^\text{es}_{n,k})^2+(Q^\text{es}_{n,k})^2}\leq \overline{S}^\text{es}_n \label{eq:P1_battery_inv_limit}
\end{align}
\end{subequations}
where $\overline{P}^\text{es}_n$ is the maximum absolute limit on charging (and discharging power), while $\overline{E}^\text{es}_n$ and $\underline{E}^\text{es}_n$ are the respective maximum and minimum limits of SoC. The storage inverter capacity is denoted by $\overline{S}^\text{es}_n$\,.

The decision variables (i.e., dispatchable quantities) considered in this work are:
\begin{align*}
    u&:=\left(
    \mathbf{P}_k^\text{pv},\,\mathbf{Q}_k^\text{pv},\mathbf{P}_k^\text{dg},\,\mathbf{Q}_k^\text{dg},\mathbf{P}_k^\text{es},\,\mathbf{Q}_k^\text{es},\mathbf{P}_k^\text{load},\,\mathbf{Q}_k^\text{load}
    \right)
\end{align*}
where the bold-faced $\mathbf{P}_k^\text{pv}=\left[P^\text{pv}_{n,k}\right]$ denotes the vector of the solar active power variables at all buses, at time $k$, and likewise for the rest. The following additional physical constraints are placed on the solar and load dispatch.
\begin{itemize}
    \item \textit{Solar curtailment}: In a predictive optimization setting, we assume the availability of solar forecast $\widehat{P}^\text{pv}_{n,k}$\,, such that the dispatched solar active power should satisfy
    \begin{align}\label{eq:pv_curt1}
        0\leq {P}^\text{pv}_{n,k}\leq \widehat{P}^\text{pv}_{n,k}\,.
    \end{align}
    Any available solar not dispatched is henceforth referred to as the solar curtailment:
    \begin{align}\label{eq:pv_curt2}
    {P}^\text{pv,curt}_{n,k}:=\widehat{P}^\text{pv}_{n,k}    -{P}^\text{pv}_{n,k}\,.
    \end{align}
    
    \item \textit{Load curtailment}: We assume that each load bus, the \textit{desired}, $P^\text{load,des}_{n,k}$\,, and \textit{minimum (critically necessary)}, $P^\text{load,min}_{n,k}$\,, load values are specified/available. Therefore the dispatched load value should follow certain limits:
    \begin{align}\label{eq:load_curt1}
        P^\text{load,min}_{n,k}\leq P^\text{load}_{n,k}\leq P^\text{load,des}_{n,k}\,.
    \end{align}
    Typically, the range of specified flexibility is narrower for critical loads for which the specified minimum value would be close to (or, same as) the desired value. Any unmet load is referred to as load curtailment:
    \begin{align}\label{eq:load_curt2}
        P^\text{load,curt}_{n,k}:=P^\text{load,des}_{n,k}-P^\text{load}_{n,k}\,.
    \end{align}
\end{itemize}
\begin{remark}
Additional constraints can also be included in the formulation. For example, constraining the power factor could be important to alleviate power quality concerns. As such, we can place constraints such as:
\begin{align*}
    \left|Q_{n,k}^\text{pv}\right|\leq \gamma^*\left|P_{n,k}^\text{pv}\right|\, ~\text{and}~ \left|Q_{n,k}^\text{es}\right|\leq \gamma^*\left|P_{n,k}^\text{es}\right|\,, 
\end{align*}
where $\gamma^*>0$ is some suitably chosen values.
\end{remark}

The objective function of the optimization problem consists of costs of generating from the DG unit, and penalties on the solar and load curtailment. In this paper, we adopt a multi-period linear cost function defined as follows:
\begin{align}
    f(u):=\sum_k \sum_n \left(c_1 P^\text{dg}_{n,k}+c_2 P^\text{pv,curt}_{n,k}+c_3 P^\text{load,curt}_{n,k}\right)
\end{align}
where the parameters $c_{1,2,3}$ are chosen appropriately to reflect the cost (or, penalty) on each term. Several works in the literature, such as~\cite{parisio2014model,zhang2013robust}, have shown that using a linear cost function in microgrids is a reasonable assumption. The baseline optimization is then given as below:
\begin{align}\label{eq:baseline}
    \textsc{(baseline)}\qquad\min~f(u)\,,~\text{s.t.}~ \text{constraints \eqref{eq:pf_linear}-\eqref{eq:load_curt2}.}
\end{align}

\section{\uppercase{Robust Optimization Problem}}
\label{sec:verification}



We adapt the optimization problem \eqref{eq:baseline} to robustify it against a specified set of adversarial events. The main idea behind robustifying the optimization dispatch is to ensure \textit{sufficient reserves} in the dispatched resources (DG, PV, storage, and interruptible load) so that these can adjust their output in response to adversarial events. While we omit the detailed description of the robust formulation due to space limitations, and refer the interested readers to \cite{liu2017robust} (and the references therein), we present here the essential deviations from the baseline formulation \eqref{eq:baseline}.
The adversarial events are grouped into the \textit{event} (or, perturbation) vector $w$\,, which may represent various cyber-physical attacks and/or physical disruptions. For example, an outage in DG could be represented by an uncertainty in the parameter $\overline{S}^\text{dg}_{n,k}$\,. Likewise, a load-masking attack could be represented by an uncertainty in $P^\text{load,des}_{n,k}$\,, while a sudden loss in solar generation could be represented by an uncertainty in $\widehat{P}^\text{pv}_{n,k}$\,. In this paper, we consider a bounded set of uncertainties:
\begin{align}
    w\in\Omega_W\,.
\end{align}
In the most general form, the robust optimization problem for a system can be expressed in the following form:
\begin{subequations}\label{eq:arbit_opt}
   \begin{align}
   & \min_{u} \max_{{w}} f({u},w)\\
    s.t. & \ g({u},{w}) \le 0 \qquad \forall {w}\in \Omega_W
\end{align}
\end{subequations}
where $g({u},{w})$ are the constraints \eqref{eq:pf_linear}-\eqref{eq:load_curt2}. The above optimization problem \eqref{eq:arbit_opt} can be converted to a robust formulation using the \textit{explicit maximization method} (see \cite{bai2015robust}) in the following form:
\begin{align}
       \min_{{u}} \widehat{f}({u})\,,\quad \text{s.t.} \,~\widehat{g}({u}) \le 0 
\end{align}
where $\widehat{f}({u})\!=\!\max_{{w}} f({u},{w})$ and $\widehat{g}({u})\!=\!\max_{{w}} g({u},{w})$. The new model does not contain any disturbance terms and if the functions $\widehat{f}({u})$ and $\widehat{g}({u})$ are convex, then the optimization problem is convex, which can be solved efficiently.

The additional decision variables we consider in our robust formulation include:
\begin{itemize}
    \item \textit{Solar reserves}: These are the up ($R^{\text{pv}+}_{n,k}>0$) and down ($R^{\text{pv}-}_{n,k}>0$) reserves set aside on the solar generation dispatch, so that the solar output has flexibility to increase or decrease from the optimally dispatched amount ($P^\text{pv}_{n,k}$) in response to an adversarial event:
    \begin{align}
        R^{\text{pv}-}_{n,k}\leq {P}^\text{pv}_{n,k}\leq \widehat{P}^\text{pv}_{n,k}-R^{\text{pv}+}_{n,k}
        \label{eq:pv_reserves}
    \end{align}
    
    \item \textit{DG reserves}: These are the up ($R^{\text{dg}+}_{n,k}>0$) and down ($R^{\text{dg}-}_{n,k}>0$) reserves set aside on the DG dispatch:
    \begin{align}
        R^{\text{dg}-}_{n,k}\leq {P}^\text{dg}_{n,k}\leq \overline{S}^\text{dg}_{n,k}-R^{\text{dg}+}_{n,k}
        \label{eq:dg_reserves}
    \end{align}
    
    \item \textit{Storage reserves}: These are the up ($R^{\text{es}+}_{n,k}>0$) and down ($R^{\text{es}-}_{n,k}>0$) reserves set aside on the storage dispatch, allocated in such a way that there is flexibility on both the charging/discharging rates, as well as the SoC limits (assuming a look-ahead period of $\Delta t$):
    \begin{subequations}
           \begin{align}
    -\overline{P}^\text{es}_{n,k}+R^{\text{es}-}_{n,k}&\leq {P}^\text{es}_{n,k}\leq \overline{P}^\text{es}_{n,k}-R^{\text{es}+}_{n,k}\\
    \underline{E}_n^\text{es} + R^{\text{es}+}_{n,k}\Delta t&\leq E_{n,k}^\text{es}\leq \overline{E}_n^\text{es}-R^{\text{es}-}_{n,k}\Delta t\,.
    \end{align}
     \label{eq:es_reserves}
    \end{subequations}
    
    \item \textit{Load reserves}: These are the up ($R^{\text{load}+}_{n,k}>0$) and down ($R^{\text{load}-}_{n,k}>0$) reserves set aside on the load dispatch (note the reverse convention on direction):
    \begin{align}
    P^\text{load,min}_{n,k}+R^{\text{load}+}_{n,k}\leq P^\text{load}_{n,k}\leq P^\text{load,des}_{n,k}-R^{\text{load}-}_{n,k}
        \label{eq:load_reserves}
    \end{align}
\end{itemize}

The goal of the robust microgrid optimization is then to minimize the cost of dispatched generation and reserves while satisfying the device and network constraints for all the disturbances in the selected disturbance set $\Omega_w$. Combining the additional reserves cost and the additional constraints \eqref{eq:pv_reserves}-\eqref{eq:load_reserves}, we develop the following robust formulation:
\begin{subequations}
\label{eq:robust}
       \begin{align}
       \textsc{(robust)}\,~\, \min_{{u}} & \,~\widehat{f}({u})+f^\text{R}(\textbf{R}^{\text{pv}}_{n,k},\textbf{R}^{\text{dg}}_{n,k},\textbf{R}^{\text{es}}_{n,k},\textbf{R}^{\text{load}}_{n,k})\\
       \text{s.t.} \,~&\,~\widehat{g}({u}) \le 0\,,\,~\text{constraints \eqref{eq:pv_reserves}-\eqref{eq:load_reserves}} 
\end{align}
\end{subequations}
where the bold-faced $\textbf{R}^{\text{pv}}_{n,k}=(R^{\text{pv}+}_{n,k},\,R^{\text{pv}-}_{n,k})$ and likewise, while $f^R$ is a linear cost function of the allocated reserves, defined similarly as $f$\,. 

Finally, we implement a simple measurement based control algorithm that dispatches reserves in response to disturbance events. In microgrids, the amount of flexible generation is increased/decreased based on real-time frequency measurements. In our quasi-static analysis, we consider the supply-demand power imbalance to be the measure of frequency and utilize that to dispatch the microgrid reserves in order to meet the demand under disturbance events. While works such as \cite{brahma2020optimal} has explored such feedback control methods, in this work, we employ a simple proportional controller that dispatches reserves in proportion to their available capacity:
\begin{align*}
    Reserve \  Dispatch=\frac{Reserve \  Capacity}{Total \  Reserves}*Imbalance\,.
\end{align*}
%



\section{\uppercase{Adversarial Set Characterization}}
\label{sec:adv_set}
While the robust formulation \eqref{eq:robust} can be solved for a given uncertainty set $\Omega_w$\,, it is not known \textit{a priori} what all adversarial events (as specified by the set) the system could be resilient against. In other words, it could be possible that, for a specified adversarial set $\Omega_w$, the problem \eqref{eq:robust} returns infeasible. In this paper, we attempt to address this fact directly by proposing an algorithmic method to characterize the (maximal) adversarial set that the system can tolerate without violation of any critical constraint.

The optimal dispatch $\mathbf{u}^\star$ obtained by solving the optimization problem \eqref{eq:pf_linear}-\eqref{eq:load_curt2} are the set-points associated with the diesel generators, dispatchable loads and solar generation. The goal of the adversarial set characterization is to gain an understanding of the set of adversarial events that can be mitigated by the reserves associated with the optimal dispatch. 
For brevity, we will denote the \textit{robust} version of the constraints \eqref{eq:pf_linear}-\eqref{eq:load_curt2} which jointly considers the decision variables $u \in \mathbb{R}^n$ and the adversarial event $w \in \Omega_W \subset \mathbb{R}^m$  as a convex constraint $g(u, w) \leq 0$.

A polytopic inner approximation $\bar{\Omega}_w$ of $\Omega_w$ that captures the set of adversarial events which can be mitigated by the reserves associated with the optimal dispatch $\mathbf{u}^\star$ can be constructed by solving $m$ optimization problems as follows:
\begin{subequations}
\begin{align}
    \mathbf{P_i:}~ &\max_{w_i}~\alpha_i \\
    &g(\mathbf{u}^\star, w_i) \leq 0\\
    &w_i \in \Omega_w\\
    &w_i = w^{nom} + \alpha_i e_i 
\end{align}
\end{subequations}
where $i \in \{1,..m\}$, $e_i \in \mathbb{R}^m$ is the i'th basis vector and $w^{nom}$ is the nominal value of the adversarial disturbance. The nominal value is simply the best  estimate of the adversarial disturbance. In the absence of any information, we can assume that $w_{nom}$ to correspond to the event where there has been no adversarial interference. The optimal solution to $\mathbf{P_i}$ is the vector $w_i^\star$ which maximizes the magnitude of the i'th component of the adversarial vector $w$. We will denote $w_0^\star = w_{nom}$. Since $g$ is convex, it follows that for any convex combination $\sum_{i=1}^m \alpha_i w^\star_i$ of the optimal solutions, $g(\mathbf{u}^\star, \sum_{i=0}^m \alpha_i w^\star_i) \leq 0$. Then, we can define:
\begin{align*}
\bar{\Omega}_w = \{w ~|~ w = \sum_{i=0}^m \alpha_i w_i^\star,~\sum_{i=0}^m \alpha_i = 1, \alpha_i >= 0\}.
\end{align*}
The set $\bar{\Omega}_w$ represents an inner approximation of the set of adversarial events which can be mitigated by the available reserves given that the optimal dispatch is \textit{fixed} to $\mathbf{u}^\star$.

\section{\uppercase{Simulation Results}}
\label{sec:results}
\subsection{Baseline Optimization: Example}
\begin{figure}[t]
\centering
\includegraphics[width=0.45\textwidth]{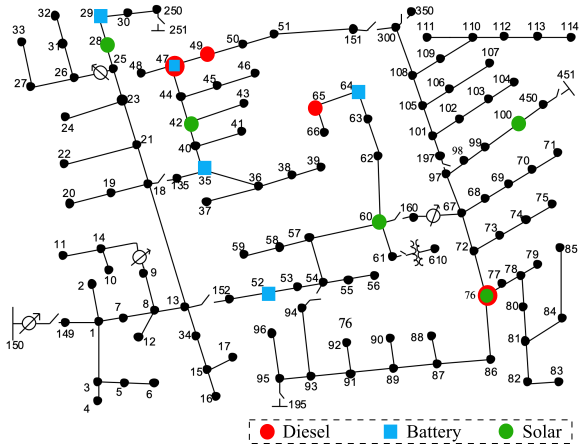}
\caption{\label{fig:IEEE_123}Modified IEEE-123 node system with added solar PV, diesel generators and battery units. Note that the bus 76 has collocated diesel and solar generator, while bus 47 has collocated diesel generator and battery.}
\end{figure}

We illustrate the baseline optimization problem \eqref{eq:baseline} on a modified three-phase IEEE-123 node system \cite{kersting2006distribution}, shown in Fig.~\ref{fig:IEEE_123}, which is converted into an islanded microgrid by disconnecting the substation from the utility. The figure shows the location of the solar PV units, diesel generator (DG) units and battery units in the network. {The peak load value in the network is 3.5\,MW (active) and 1.9\,MVAR (reactive). The cumulative maximum generation limit of the DGs is kept at 2.5\,MW, while the cumulative maximum PV generation is 1.77\,MW. The aggregated (charing/discharging) power limit of the batteries is 1.5\,MW, with an aggregated energy capacity of 6\,MWh.} The network model is hosted in GridLAB-D \cite{chassin2008gridlab} and interacts with the baseline optimization (implemented in Julia) via FNCS which is a co-simulation platform  \cite{ciraci2014fncs}. To showcase the baseline optimization problem, we simulate the model under two scenarios: 1) \textit{low solar high load}, and 2) \textit{high solar low load}. 

\subsubsection{Low solar high load case}
\begin{figure}[t]
\centering
\includegraphics[width=0.3\textwidth]{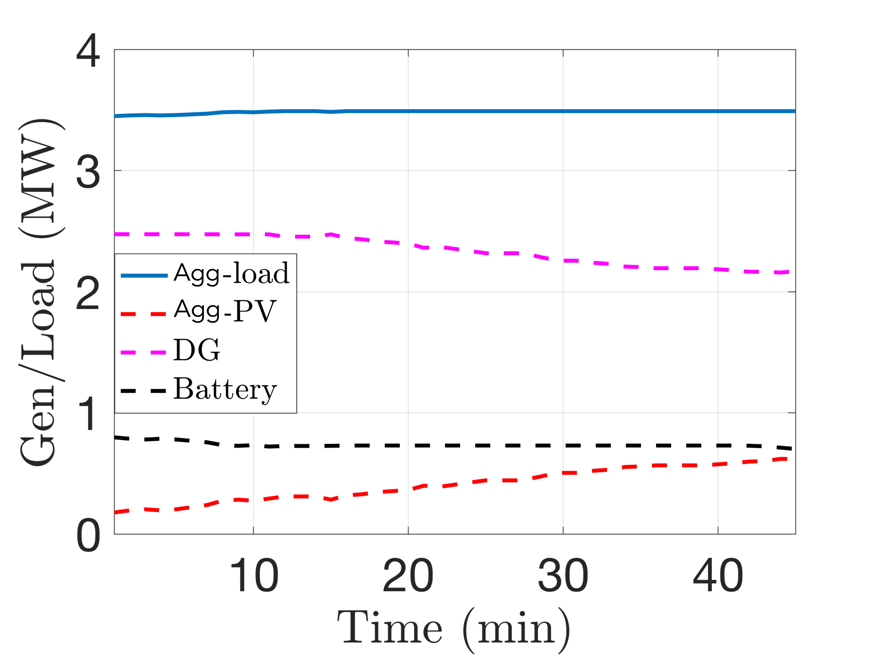}
\caption{\label{fig:Gen_load_LSHL} Generation and load in the low solar high load case.}
\end{figure}

\begin{figure}[t]
    \vspace{-9pt}
  \subfloat [\label{fig:solar_curt_LSHL}]{   \includegraphics[width=0.45\linewidth]{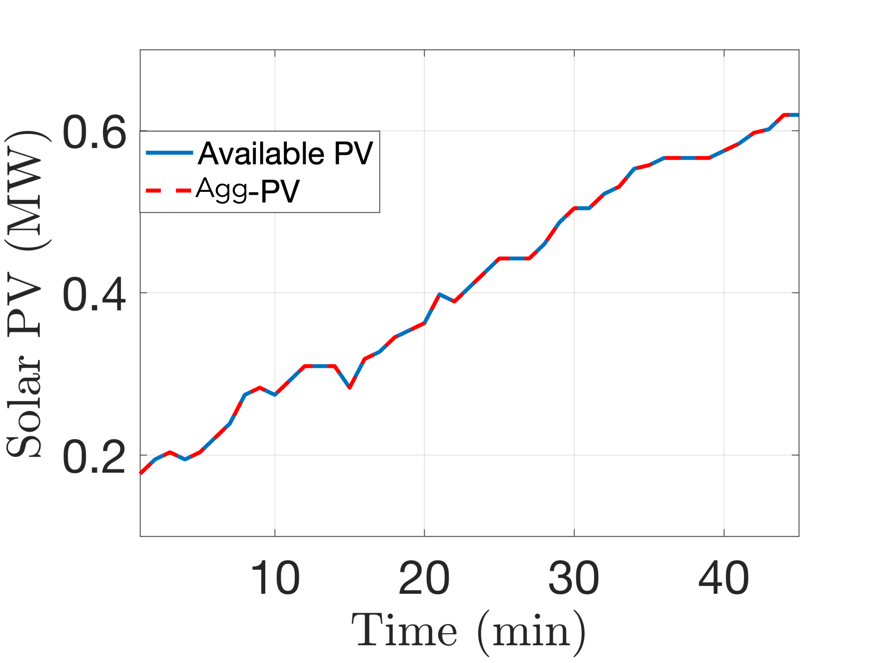}}
    \hfill
  \subfloat [\label{fig:load_curt_LSHL}]{    \includegraphics[width=0.45\linewidth]{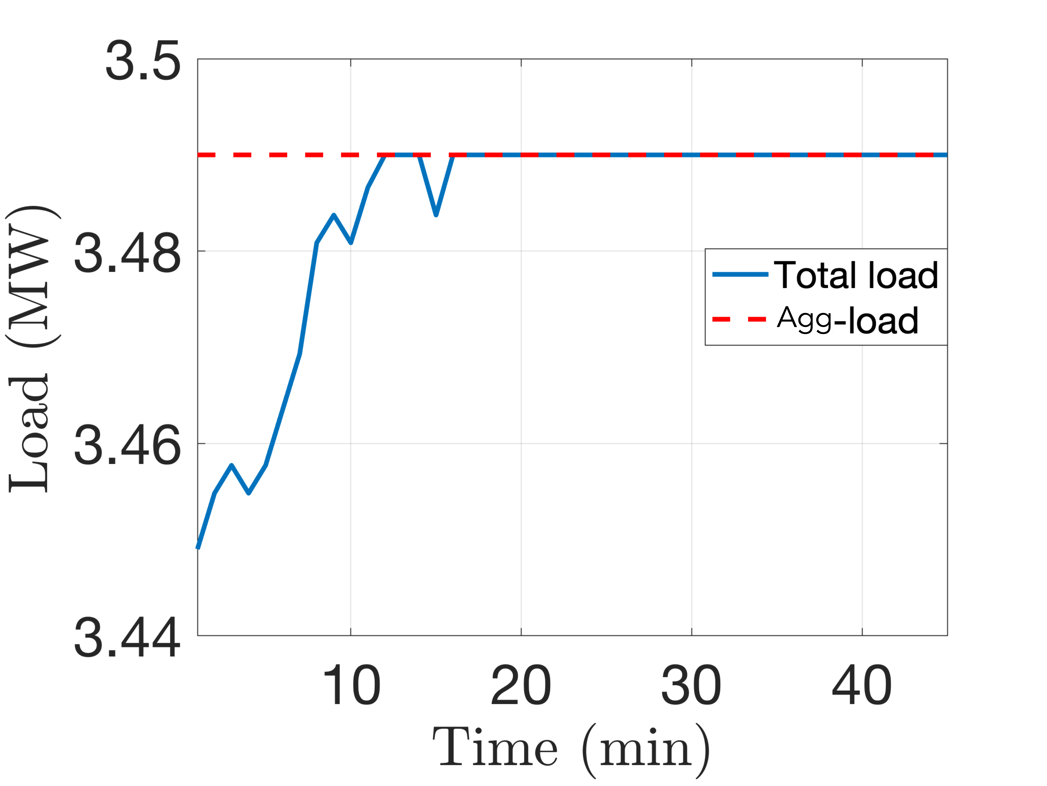}}
  \caption{Low solar high load case: (a) Solar curtailment (b) Load curtailment}
  \label{fig:LSHL_curt}
\end{figure}

\begin{figure}[t]
    \vspace{-9pt}
  \subfloat [\label{fig:volt_LSHL}]{   \includegraphics[width=0.45\linewidth]{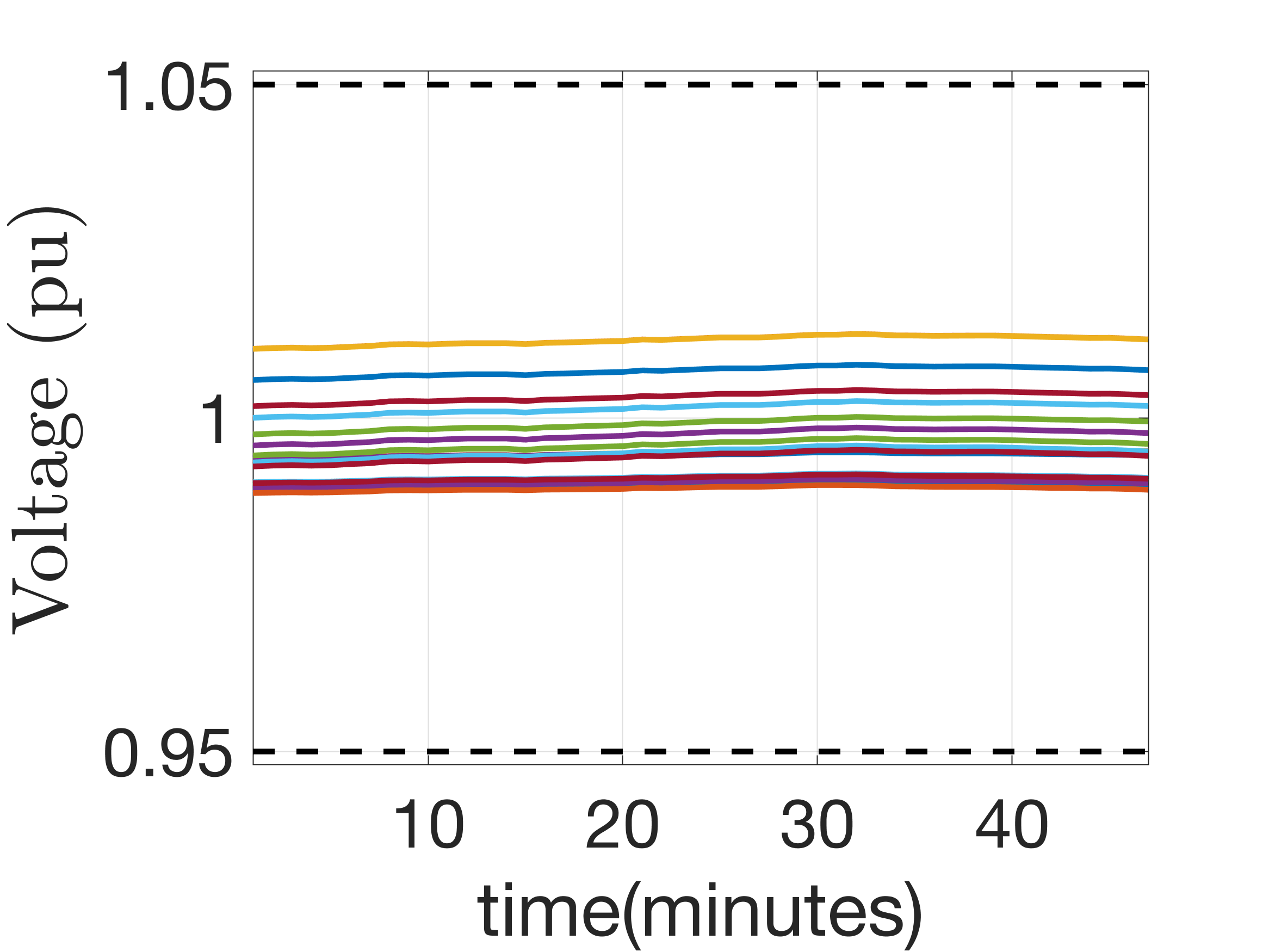}}
    \hfill
  \subfloat [\label{fig:SoC_LSHL}]{    \includegraphics[width=0.45\linewidth]{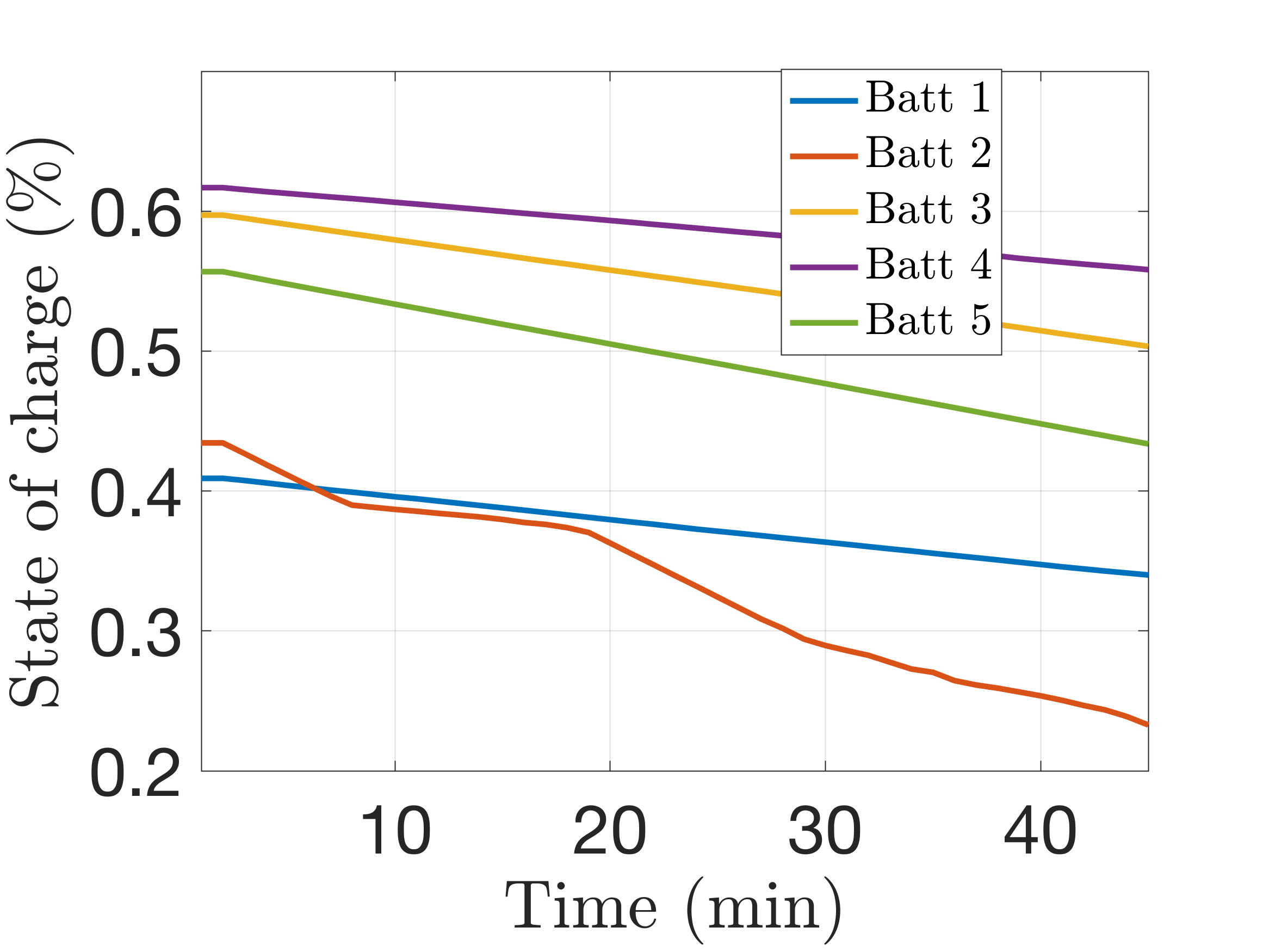}}
  \caption{Low solar high load case: (a) Nodal voltages (b) Battery state of charge evolution.}
  \label{fig:LSHL_output}
\end{figure}

This scenario represents the solar and load profiles typically observed during the morning hours, with a low but increasing amount of solar and a high load. The generation and load dispatch for this case is shown in Fig.~\ref{fig:Gen_load_LSHL} which shows how with increasing solar generation over time, the DG generation is reduced in order to reduce the cost of operation. Furthermore, Fig.~\ref{fig:LSHL_curt} shows the curtailment in solar and load, highlighting that initially there a small amount of load curtailment which is required due to inadequate solar PV available, but as the solar PV generation increases, the load curtailment becomes zero. The effect of the generation and load dispatches on the system constraints is depicted in Fig.~\ref{fig:LSHL_output} which shows that the voltages are within their operational limits and the state of charge (SoC) of each battery in the system is also within the SoC limits. In these simulation results, we assume that the initial SoC of the batteries is randomly chosen.

\subsubsection{High solar low load case} 
\begin{figure}[t]
\centering
\includegraphics[width=0.3\textwidth]{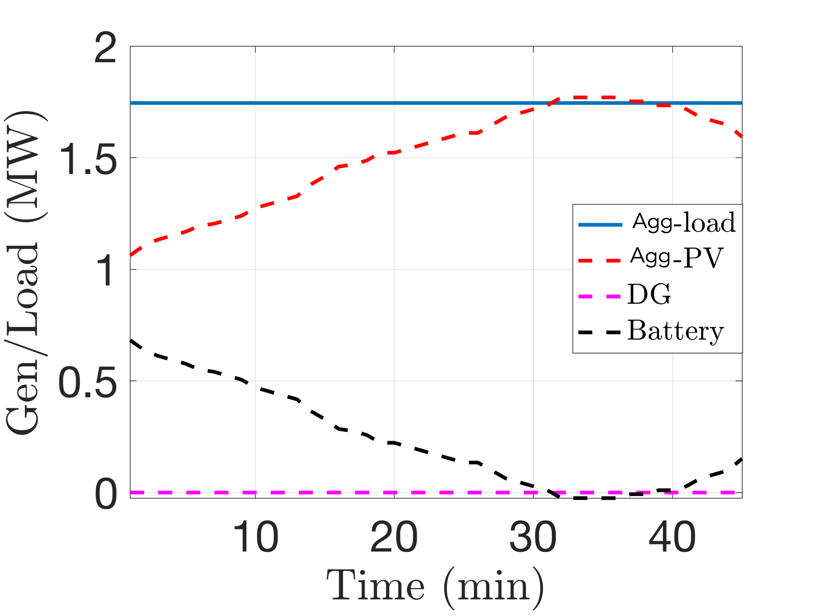}
\caption{\label{fig:Gen_load_HSLL} Generation and load in the high solar low load case.}
\end{figure}

\begin{figure}[t]
    \vspace{-9pt}
  \subfloat [\label{fig:volt_HSLL}]{   \includegraphics[width=0.45\linewidth]{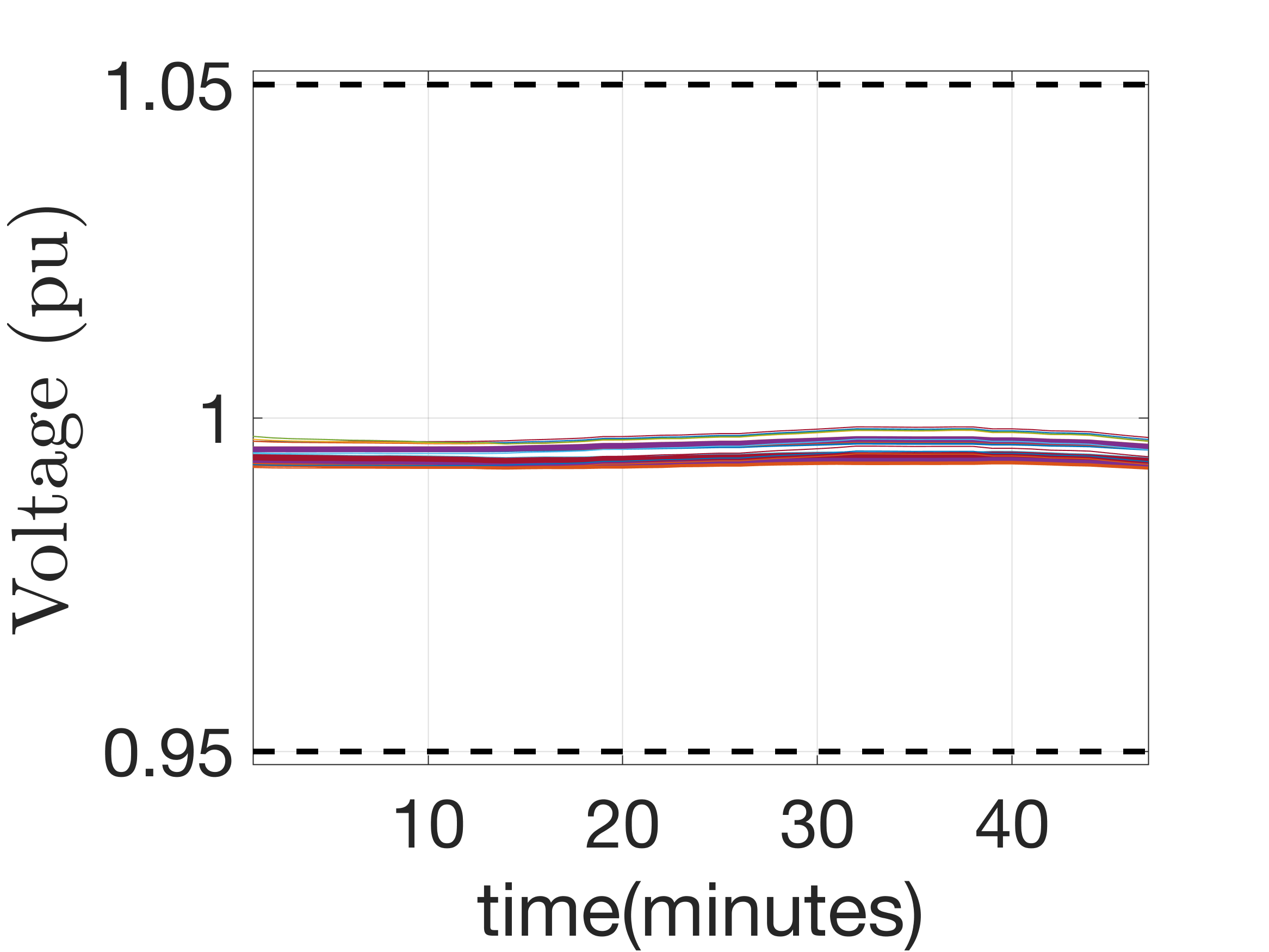}}
    \hfill
  \subfloat [\label{fig:SoC_HSLL}]{    \includegraphics[width=0.45\linewidth]{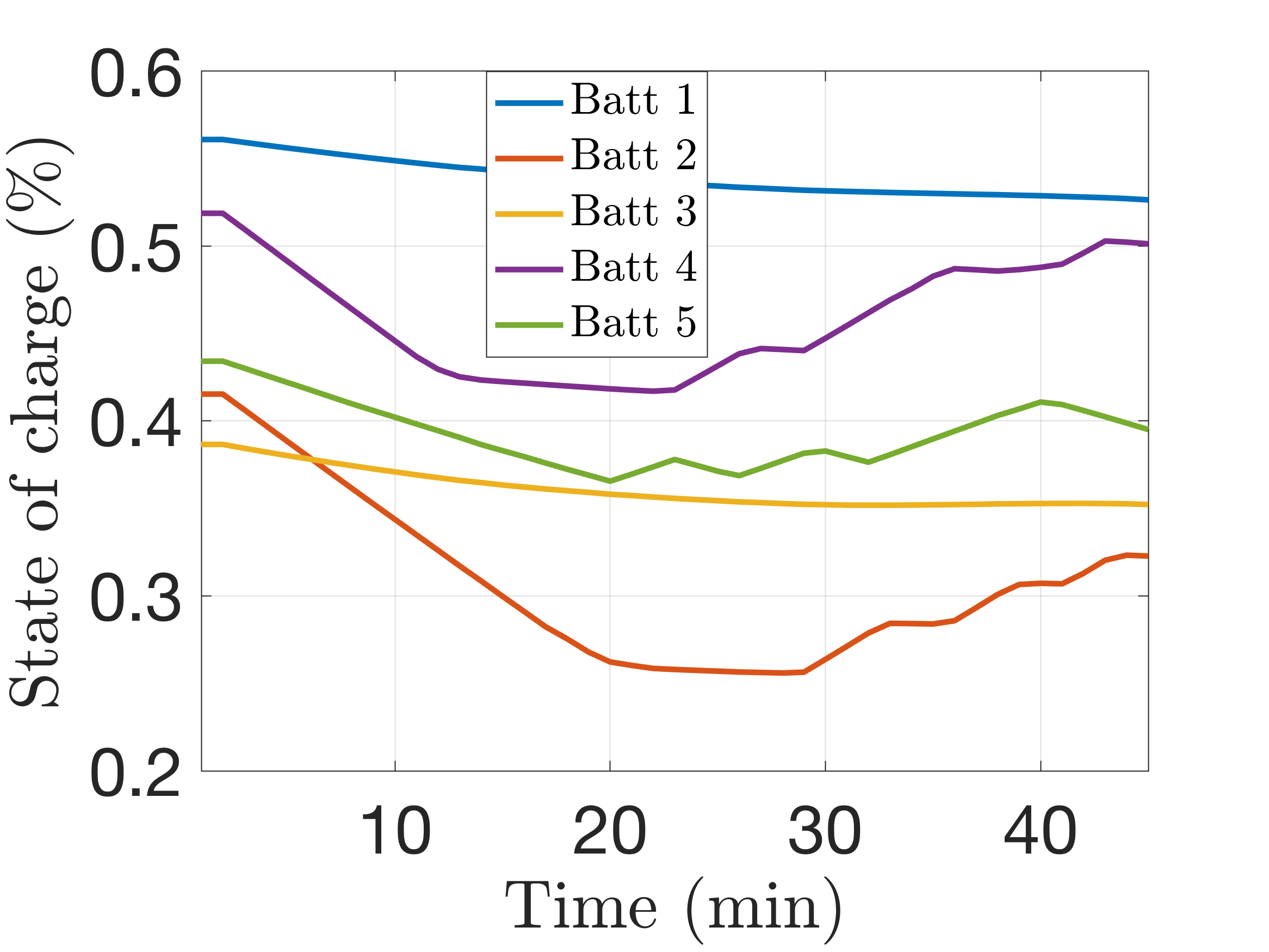}}
  \caption{High solar low load case: (a) Nodal voltages (b) Battery state of charge evolution}
  \label{fig:HSLL_output}
\end{figure}

This scenario represents the solar and load profiles typically observed during the noon/midday hours, with high solar generation but low load. The generation and load dispatch for this case is depicted in Fig.~\ref{fig:Gen_load_HSLL}, which shows how the battery generation reduces and the solar generation increases (since reducing solar curtailment is part of the objective). Also, it can be seen that the DG output is zero as there is sufficient output available from the solar generation and batteries to meet the load. The effect on the system constraints of this generation dispatch is shown in Fig.~\ref{fig:HSLL_output}, which depicts that the voltages are within their nodal limits and that the state of charge (SoC) for all the batteries is within their respective limits (in per unit).

 These case studies show that the baseline optimization satisfies supply-demand balance and is able to satisfy the voltage and state of charge constraints. Furthermore, the amount of load curtailment is minimum and supply is maintained to critical loads in the system.

 \subsection{Robust optimization simulation results}

\begin{figure}[t]
    \vspace{-9pt}
  \subfloat [\label{fig:CaseCP_DG_trip}]{   \includegraphics[width=0.45\linewidth]{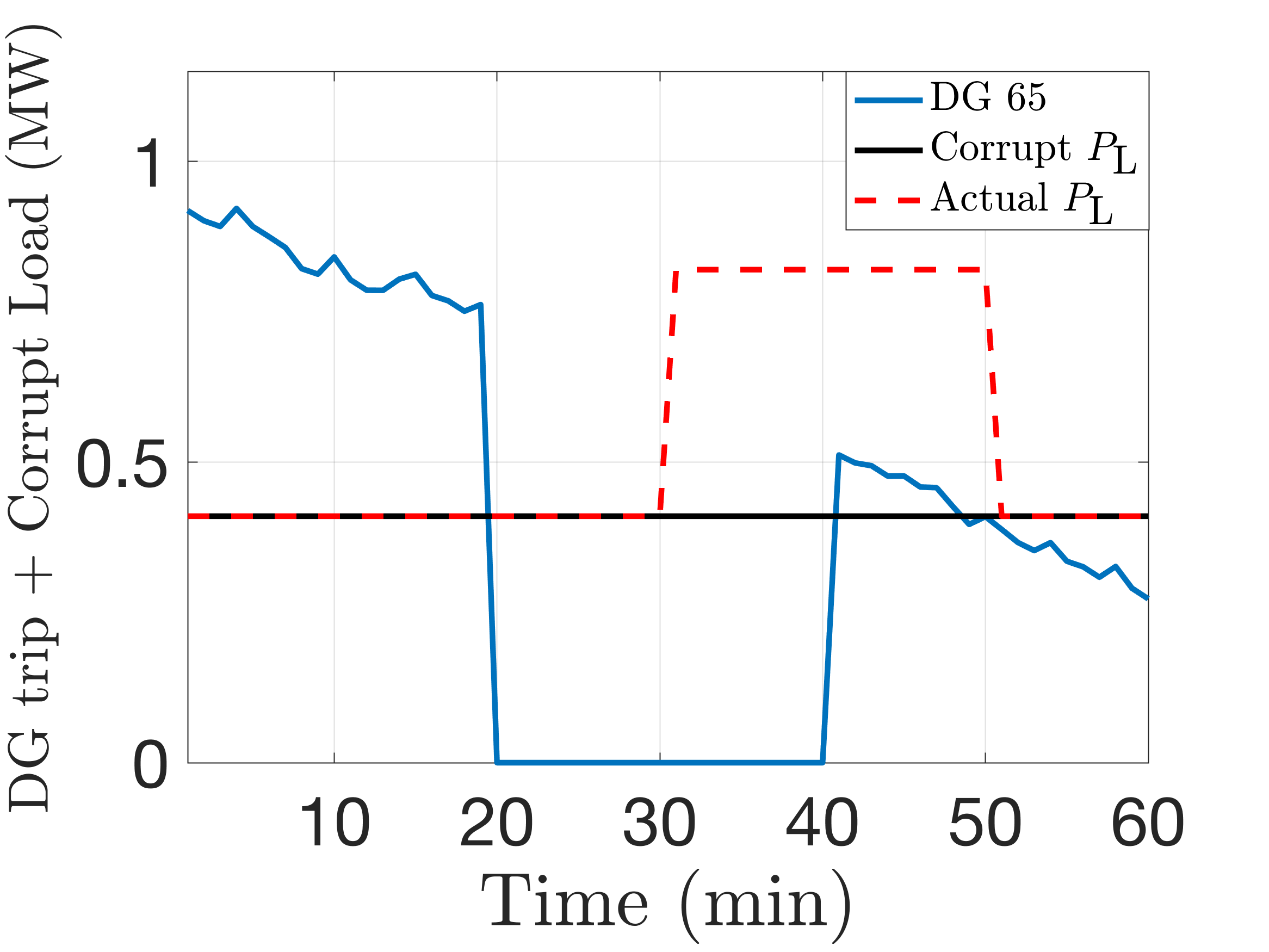}}
    \hfill
  \subfloat [\label{fig:CaseCP_agg_gen}]{    \includegraphics[width=0.45\linewidth]{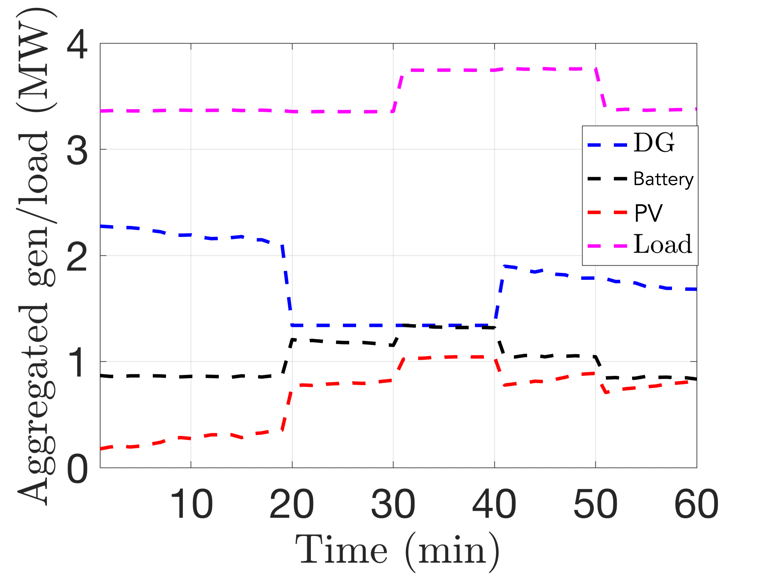}}
  \caption{Cyber-physical event case: (a) DG-trip followed by cyber-attack that masks the increase in load (b) Response of the flexible generation and curtailment to deal with this event}
  \label{fig:Case_CP_event}
\end{figure}

\begin{figure}[t]
\centering
\includegraphics[width=0.3\textwidth]{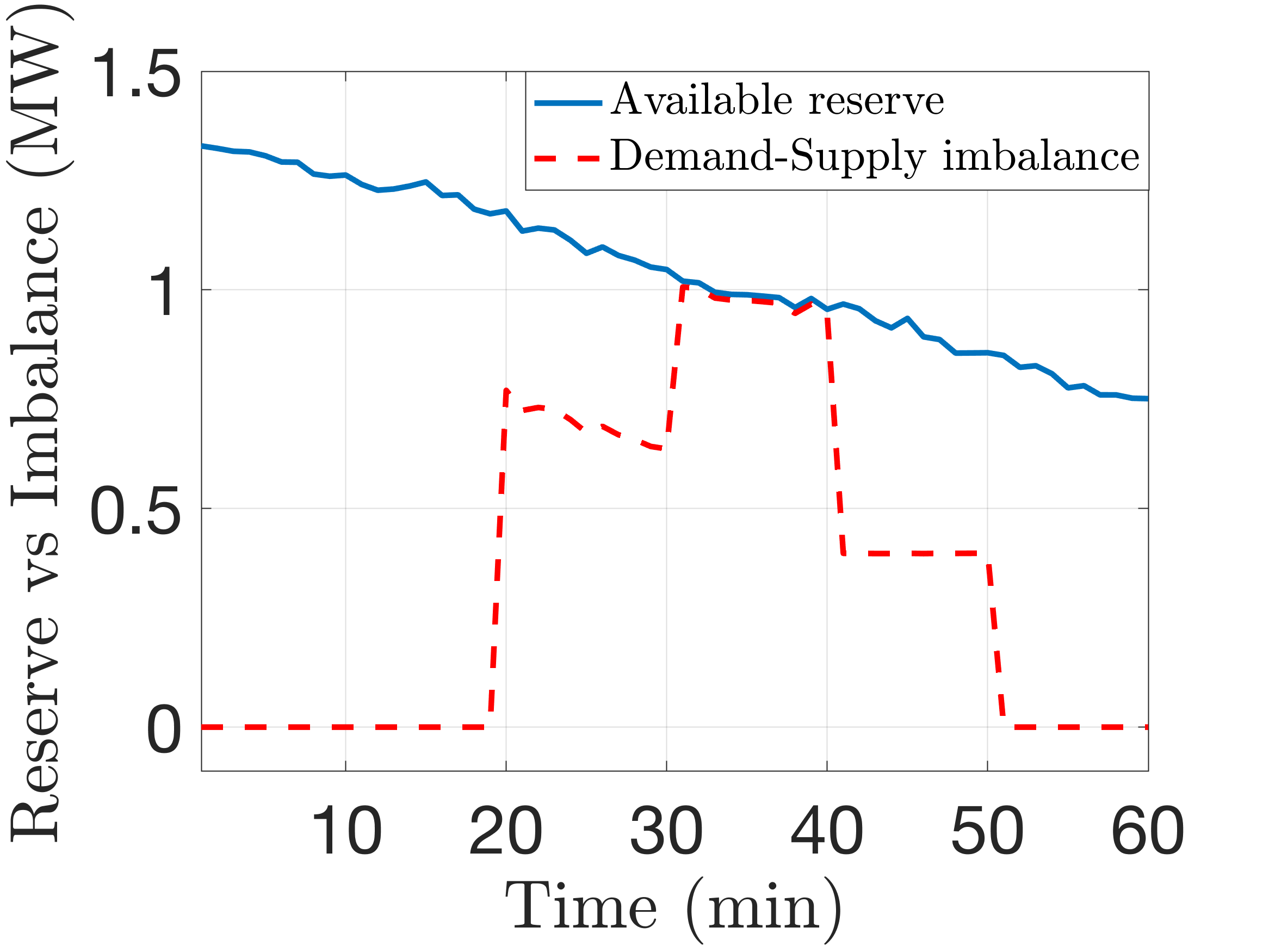}
\caption{\label{fig:CaseCP_reserve}Supply demand imbalance created by the cyber-physical event and the reserves available to deal with such an event. The figure shows that enough reserves are available to the system operator to effectively deal with this event.}
\end{figure}

\begin{figure}[t]
    \vspace{-9pt}
  \subfloat [\label{fig:CaseCP_SoC1}]{   \includegraphics[width=0.45\linewidth]{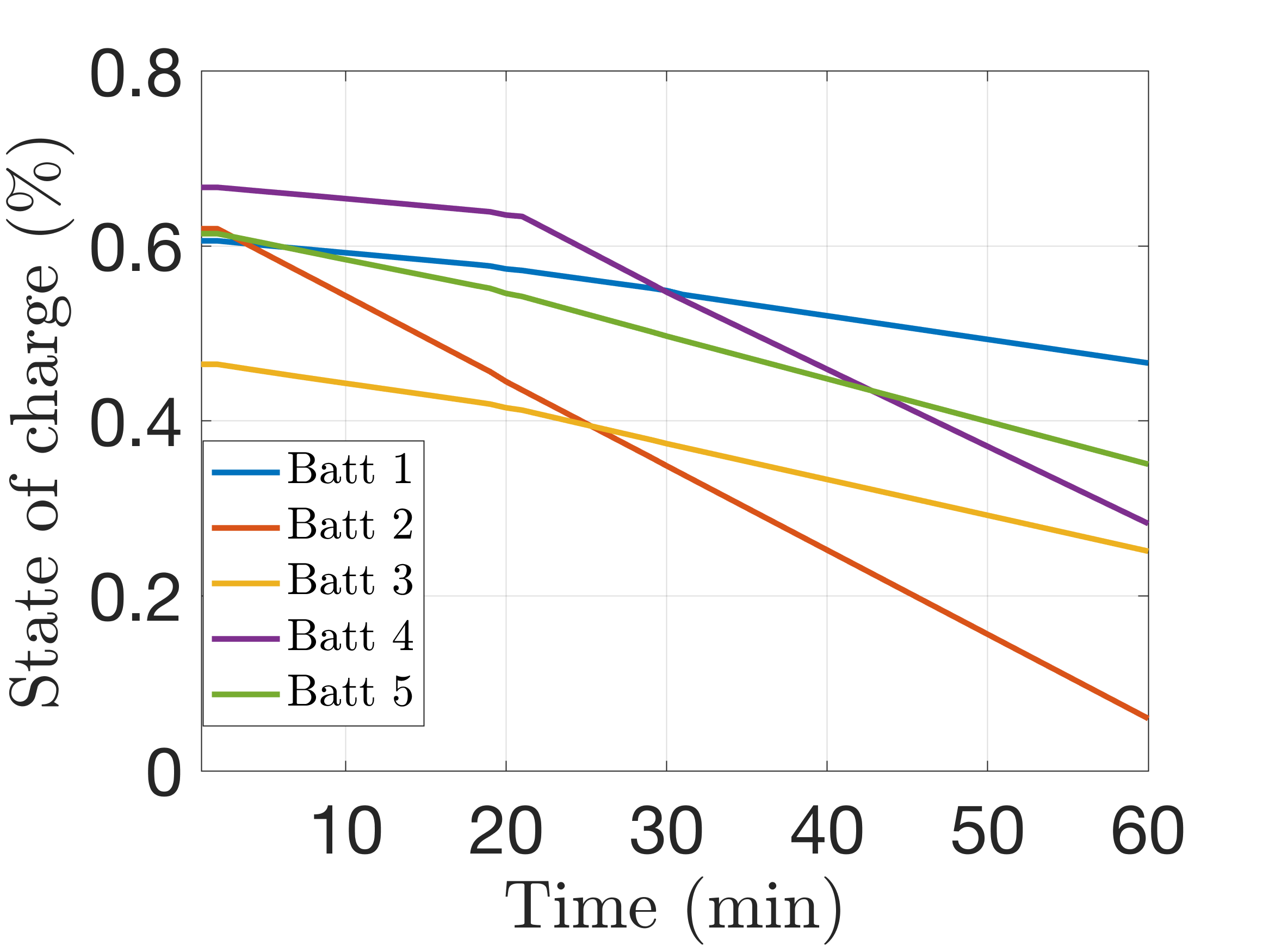}}
    \hfill
  \subfloat [\label{fig:CaseCP_volt}]{    \includegraphics[width=0.45\linewidth]{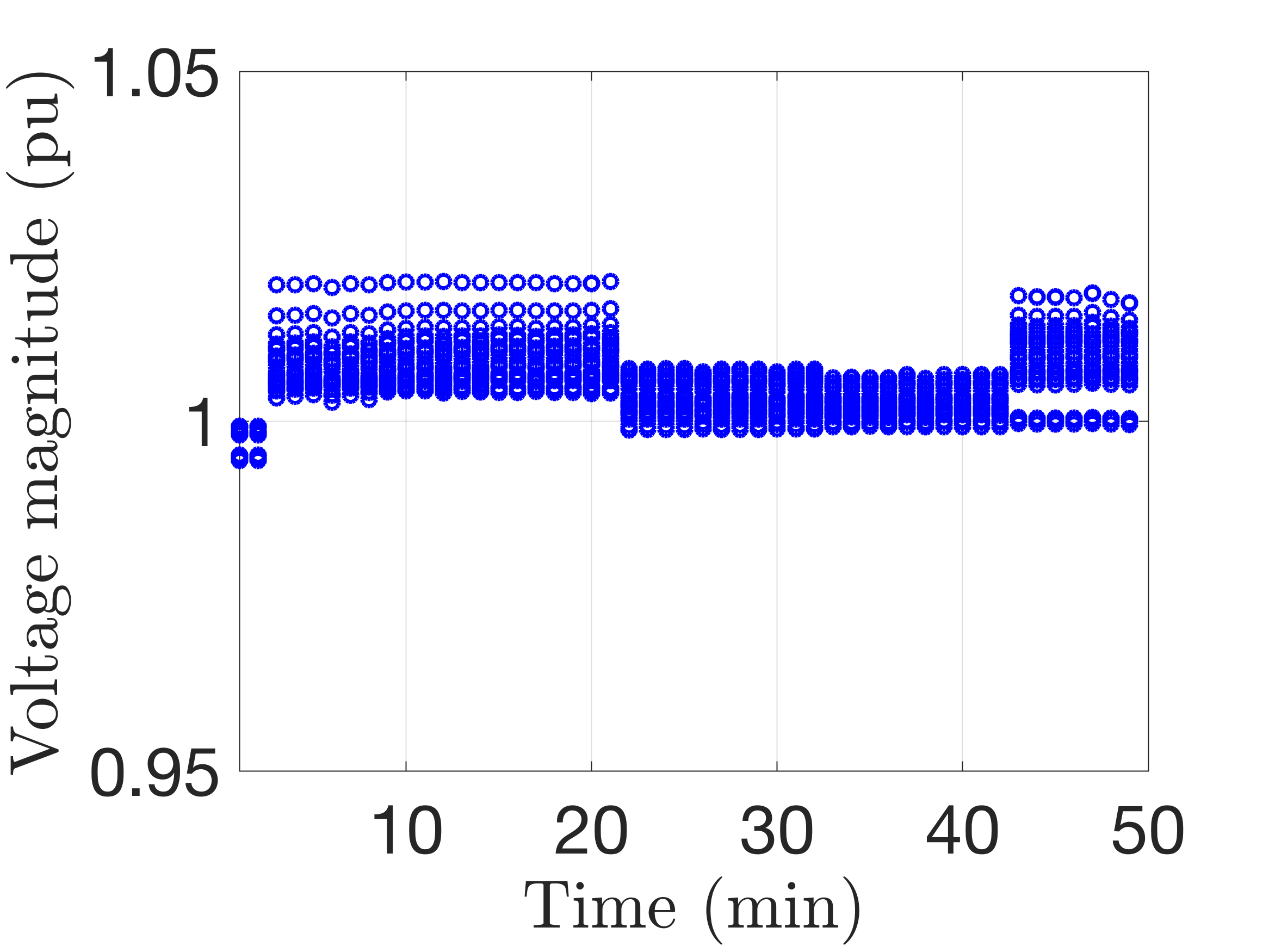}}
  \caption{Network constraint under cyber-physical event: (a) Battery state of charge (SoC) (b) Network voltages}
  \label{fig:CaseCP_network}
\end{figure}

In the following set of simulation results, we will illustrate the effectiveness of the robust formulation (in Sec.\,\ref{sec:verification}). We consider a disturbance event consisting of a cyber-physical event as depicted in Fig.~\ref{fig:CaseCP_DG_trip}, wherein the largest DG in the microgrid (DG 65) trips at time 20\,min and become offline and soon after, at 30\,min, there is a cyber attack which masks the increase in load from the system operator. The DG is made operational and comes back online at 40\,min, whereas the cyber attack is finally over at 50\,min. We solve the robust optimization problem to maintain sufficient reserves in order to meet the increased demand and also maintain system constraints under this cyber-physical event, i.e., the robust optimization problem should have enough reserves to dispatch in case such a cyber-physical event happens on the microgrid. {It is assumed that the disturbance can be measured through the imbalance in supply and demand which manifests itself in the form of frequency deviation in the system. This frequency measurement is utilized by the flexible resources to dispatch their reserves when an adversarial event occurs.} The results for the robust optimization dispatch are shown in Fig.~\ref{fig:CaseCP_agg_gen}. The results show that as the DG trips at 20\,min, the solar PV and the battery increase their output in order meet the load. Furthermore, when the cyber-attack at 30\,min masks the increased load, the battery and PV generation is further increased to meet this load. As the DG unit comes back online and the cyber-attack ends, the battery and solar PV also reduce their output. These simulation results illustrate the effectiveness of the robust formulation to deal with this cyber-physical event. Furthermore, Fig.~\ref{fig:CaseCP_reserve} shows that the robust formulation allocates enough total reserves to deal with this cyber-physical event, i.e., to counter the loss of the largest DG and the load masking cyber-attack. The results in Fig.~\ref{fig:CaseCP_network} demonstrate the effect of this robust dispatch on the system constraints, i.e., the SoC of all the batteries is within their limits and the nodal voltages are within the specified bounds.

\subsection{Adversarial set simulation results}
\begin{figure}[t]
    \vspace{-9pt}
    \subfloat [\label{fig:perturbations}]{   \includegraphics[width=0.45\linewidth]{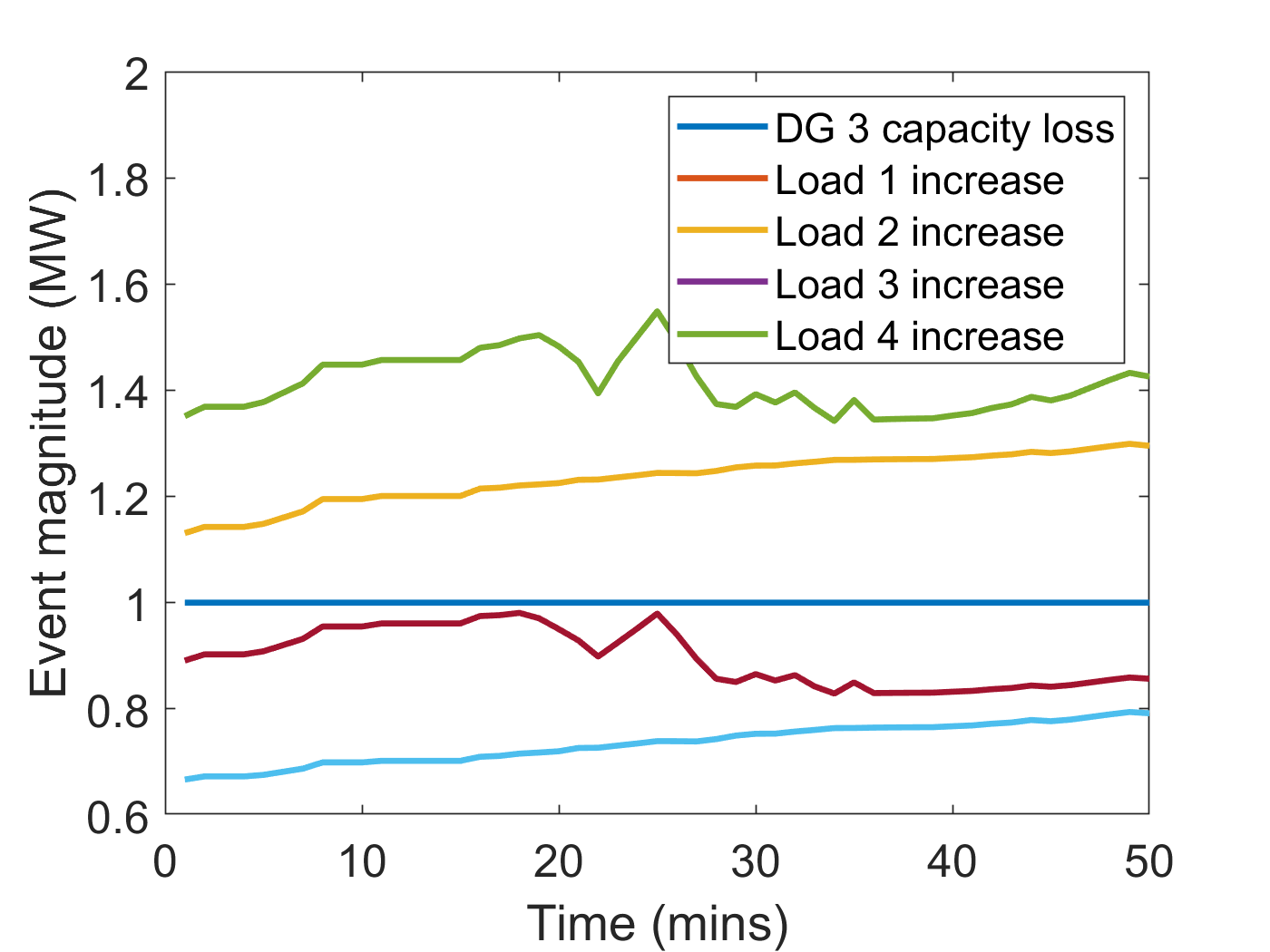}}
    \hfill
  \subfloat[\label{fig:load_region}]{
\includegraphics[width=0.45\linewidth]{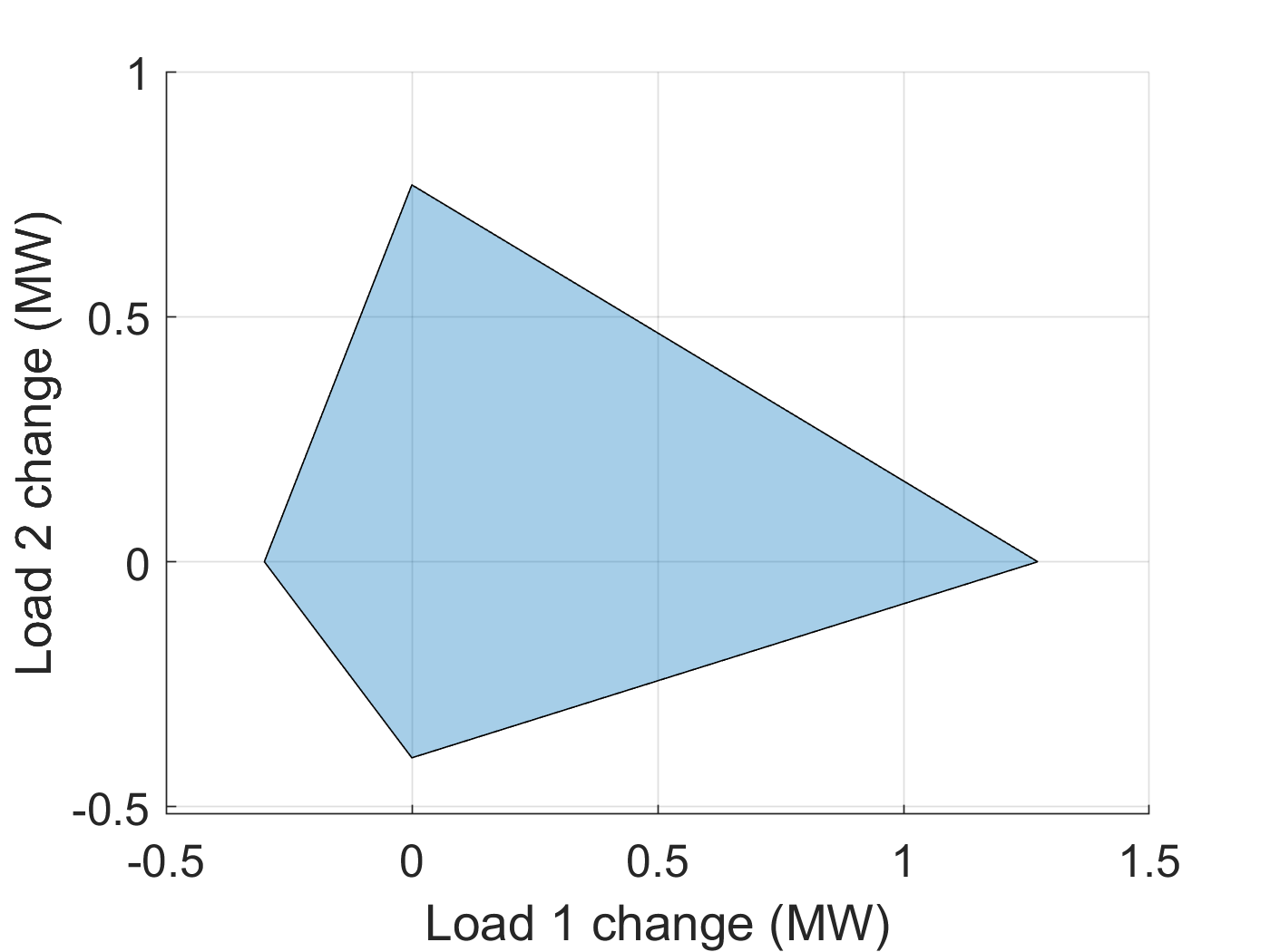}}
\caption{Adversarial set characterization: (a) The maximum component-wise perturbation $\alpha_i^\star$ that is supported by the available reserves (b) A 2-dimensional subset of $\bar{\Omega}_w$ for a specific time instant that illustrates the convex region of the tolerable variations in load 1 and 2.}
\end{figure}

\begin{figure}[t]
    \vspace{-9pt}
    \subfloat [\label{fig:pv1}]{   \includegraphics[width=0.45\linewidth]{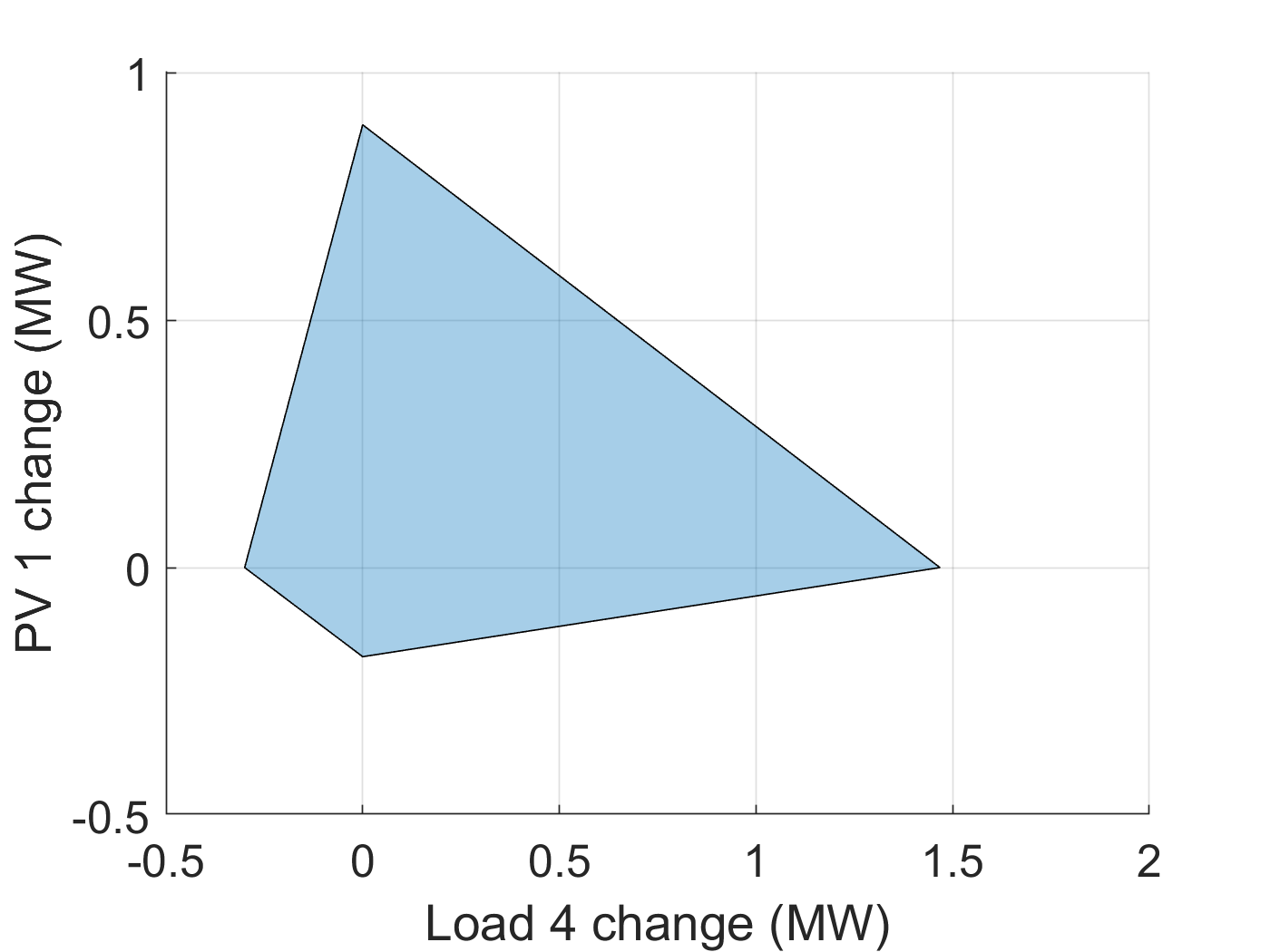}}
    \hfill
  \subfloat[\label{fig:pv2}]{
\includegraphics[width=0.45\linewidth]{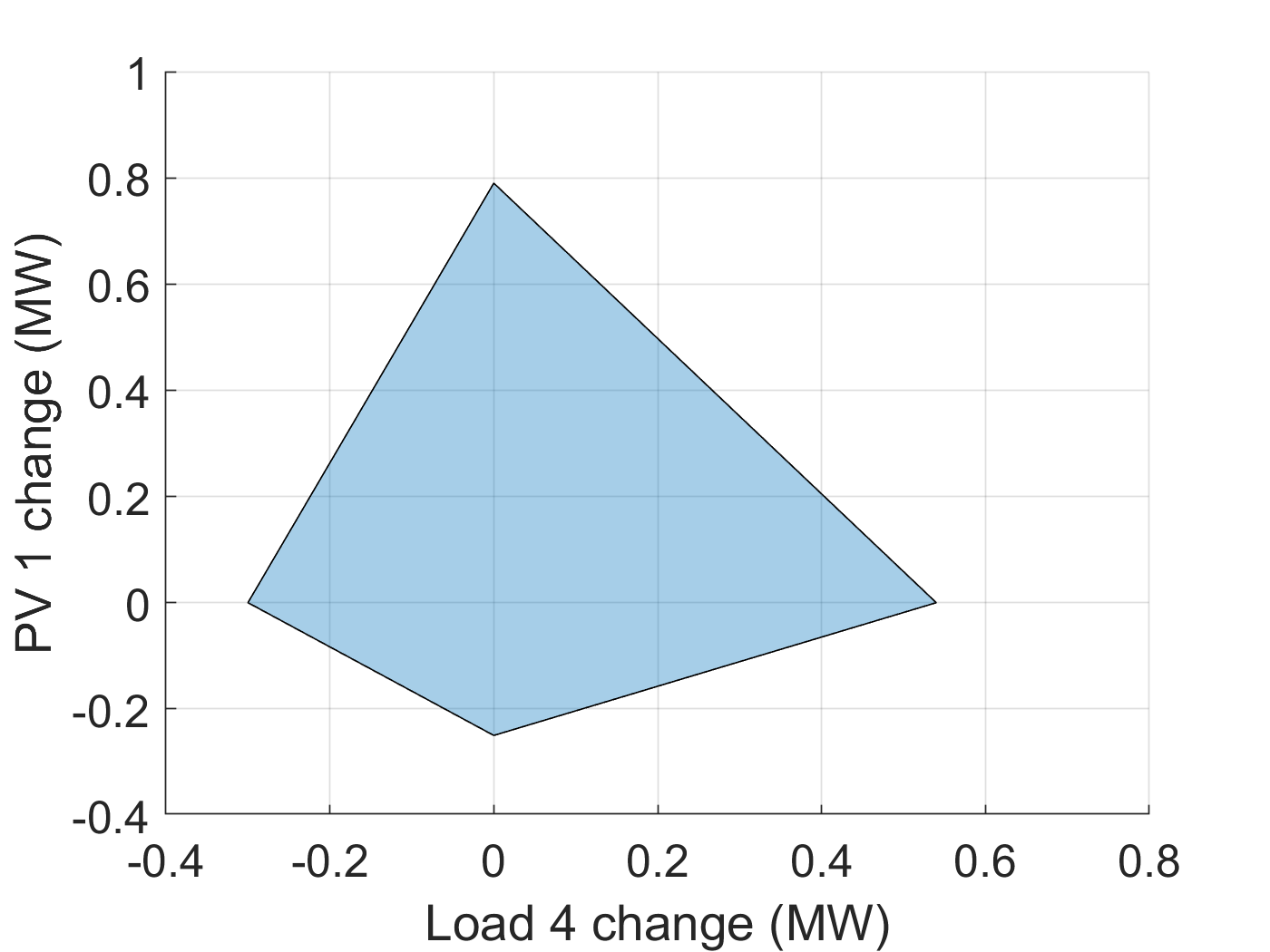}}
\caption{2-dimensional subset of the time-varying adversarial set $\bar{\Omega}_w$ capturing the allowable variations in solar PV output and load 4, at a) T = 1 minute b) T= 25 min.}
\end{figure}

\begin{figure}[t]
    \vspace{-9pt}
  \subfloat [\label{fig:CaseCP_Gen}]{   \includegraphics[width=0.45\linewidth]{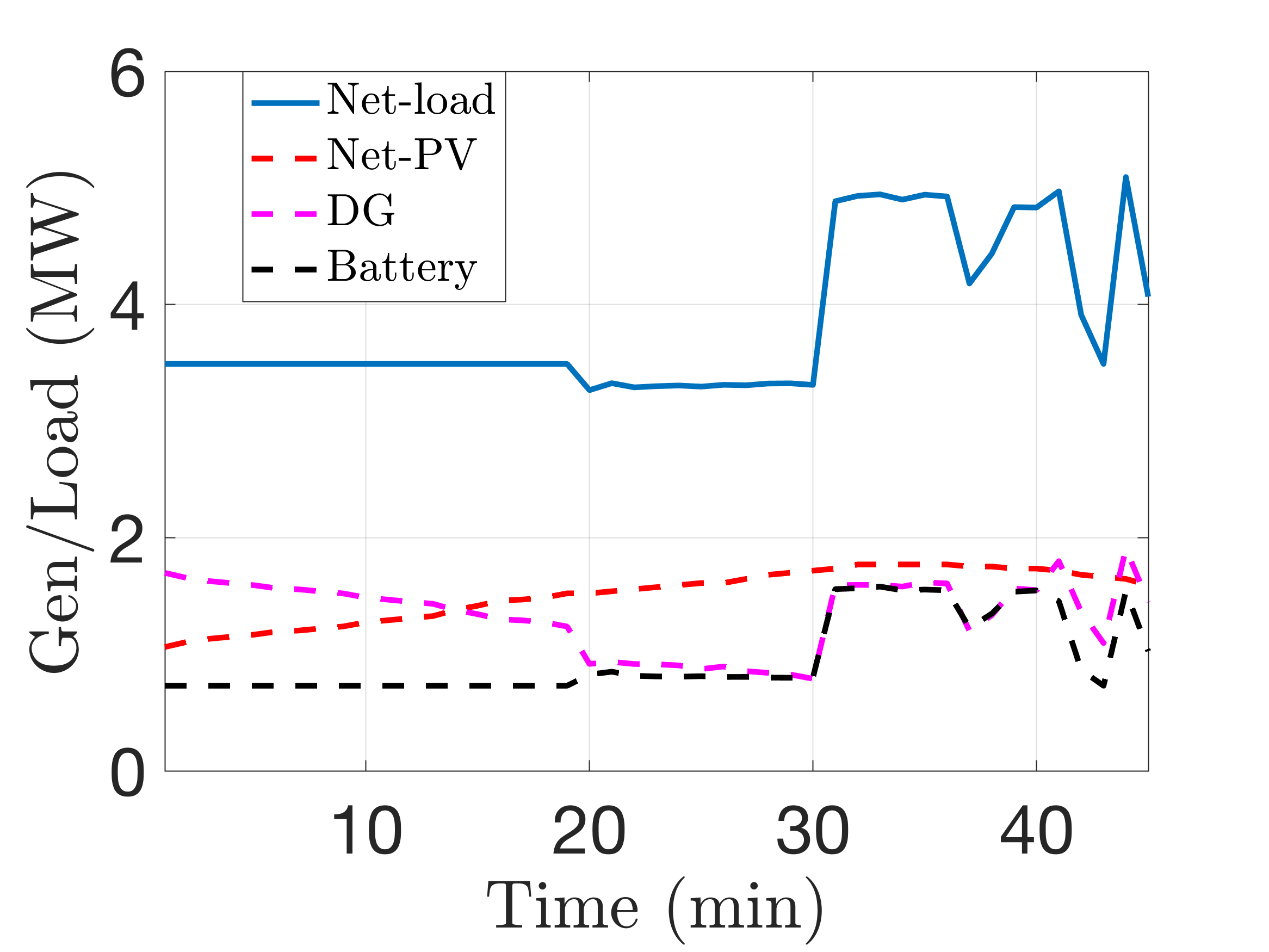}}
    \hfill
  \subfloat [\label{fig:CaseCP_SoC}]{    \includegraphics[width=0.45\linewidth]{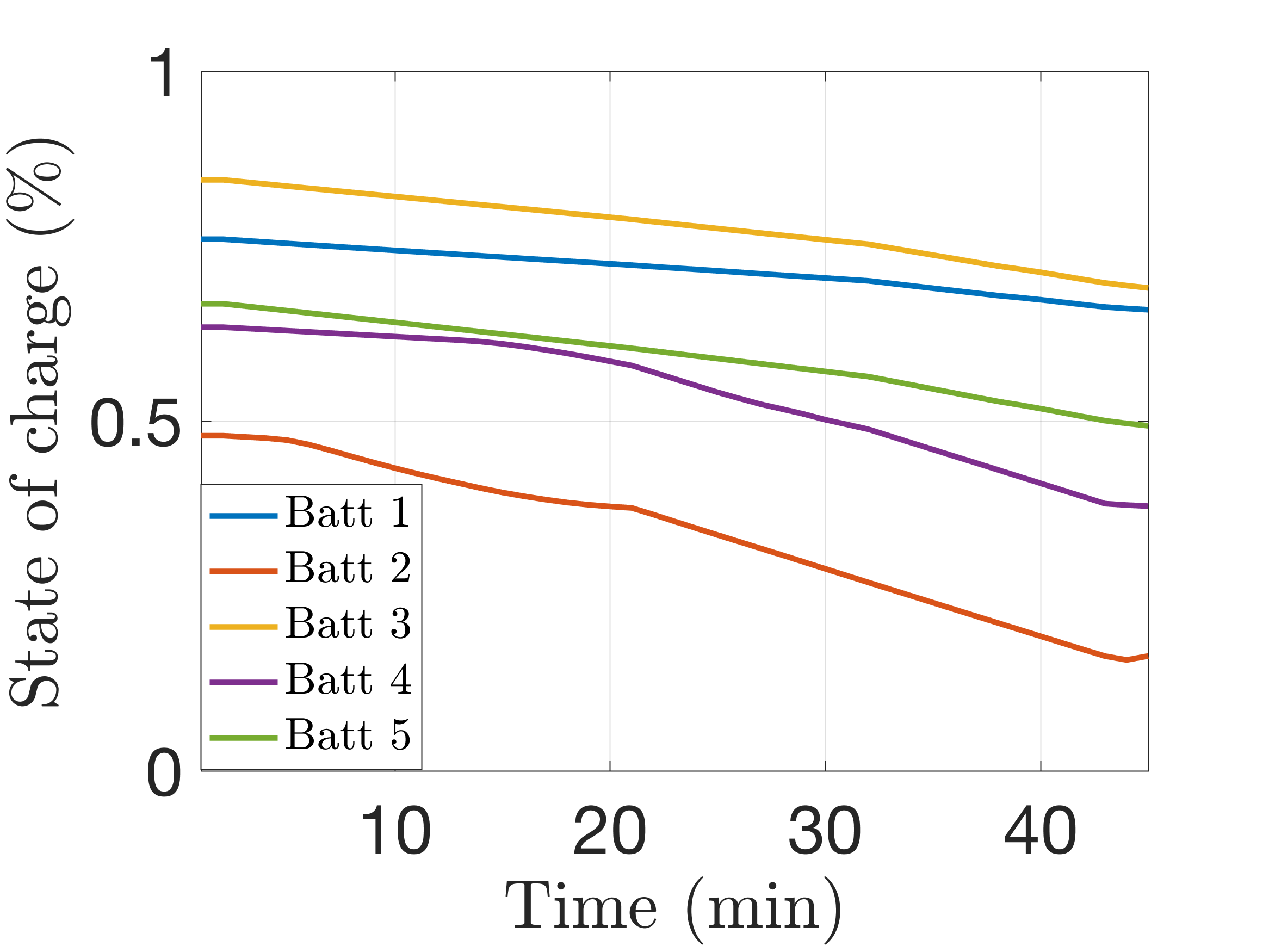}}
  \caption{Network constraint under cyber-physical event:(a) Response of the flexible generation under cyber-physical event (b) Battery state of charge (SoC)}
  \label{fig:CaseCP_Advset}
\end{figure}

In order to showcase the adversarial set characterization (presented in Sec.\,\ref{sec:adv_set}), we determine the set of largest disturbances which the system can handle. In this scenario we consider losing the largest DG unit, load increase at certain buses, and error in the solar forecast. The baseline optimization problem is solved for each time instant, and given the baseline dispatch, the set $\bar{\Omega}_w$ is computed for each time step by solving $\mathbf{P}_i$ for each potential adversarial event. The maximum component-wise perturbation that can be tolerated and a 2-dimensional projection of $\bar{\Omega}_w$ for specific time instants in shown in Fig \ref{fig:perturbations} and Fig \ref{fig:load_region}. Furthermore, the shape of the set $\bar{\Omega}_w$ changes over time and depends on the amount of flexibility that is available in the system (see Fig \ref{fig:pv1} and Fig \ref{fig:pv2}). Then sampling from the set $\bar{\Omega}_w$ showcases that these disturbances can be handled by the available flexibility. This is shown in Fig.~\ref{fig:CaseCP_Advset} which shows the response of the generators to the loss of DG unit and increase in load and also shows that in this response the SoC constraints are still satisfied. Figure~\ref{fig:CaseCP_Gen} shows that at time $T=20$ minutes there is a loss of DG (chosen randomly between zero and maximum DG loss calculated for the adversarial set). Based on the above analysis, we are guaranteed to have enough reserves available in the system to deal with this disturbance, which can be seen from the increase in battery output and solar PV generation. Furthermore, at $T=30$ minutes, we experience a load increase at certain buses (again chosen randomly at each step between zero and the maximum calculated based on the adversarial set). To deal with this load increase, we dispatch additional flexibility from the battery and the DG units. These simulation results illustrate that the maximum adversarial set calculated in the previous analysis can be managed feasibly by the available flexibility in the microgrid, while maintaining the supply-demand balance and other constraints such as SoC as depicted in Fig.~\ref{fig:CaseCP_SoC}.

\section{CONCLUSIONS}
\label{sec:concl}
This paper presented a method to determine the adversarial set of a three-phase microgrid system. We formulate the robust optimization problem and from that we formulate the adversarial set characterization problem. We illustrate the inner approximation of the adversarial set through simulation results on IEEE-123 node distribution system. Future work will develop methods to incorporate the determined adversarial sets in microgrid planning and operation. Future work will also focus on probabilistic uncertainties (in the form of chance-constrained and risk based optimization), and compare the conservativeness of the characterized adversarial set between the deterministic and probabilistic approaches. Moreover, we will also investigate the applicability of this approach to larger systems (e.g., networked microgrids).






\section*{ACKNOWLEDGMENT}

This research was supported by the ``Resilience through Data-driven Intelligently-Designed Control'' Initiative, under the Laboratory Directed Research and Development Program at Pacific Northwest National Laboratory (PNNL). PNNL is a multi-program national laboratory operated for the U.S. Department of Energy by Battelle Memorial Institute under Contract No. DE-AC05-76RL01830.



\bibliographystyle{IEEEtran}
\bibliography{References.bib}

\begin{thebibliography}{10}
\providecommand{\url}[1]{#1}
\csname url@samestyle\endcsname
\providecommand{\newblock}{\relax}
\providecommand{\bibinfo}[2]{#2}
\providecommand{\BIBentrySTDinterwordspacing}{\spaceskip=0pt\relax}
\providecommand{\BIBentryALTinterwordstretchfactor}{4}
\providecommand{\BIBentryALTinterwordspacing}{\spaceskip=\fontdimen2\font plus
\BIBentryALTinterwordstretchfactor\fontdimen3\font minus
  \fontdimen4\font\relax}
\providecommand{\BIBforeignlanguage}[2]{{%
\expandafter\ifx\csname l@#1\endcsname\relax
\typeout{** WARNING: IEEEtran.bst: No hyphenation pattern has been}%
\typeout{** loaded for the language `#1'. Using the pattern for}%
\typeout{** the default language instead.}%
\else
\language=\csname l@#1\endcsname
\fi
#2}}
\providecommand{\BIBdecl}{\relax}
\BIBdecl

\bibitem{di2018decarbonization}
M.~L. Di~Silvestre, S.~Favuzza, E.~R. Sanseverino, and G.~Zizzo, ``How
  decarbonization, digitalization and decentralization are changing key power
  infrastructures,'' \emph{Renewable and Sustainable Energy Reviews}, vol.~93,
  pp. 483--498, 2018.

\bibitem{farrokhabadi2019microgrid}
M.~Farrokhabadi, C.~A. Ca{\~n}izares, J.~W. Simpson-Porco, E.~Nasr, L.~Fan,
  P.~A. Mendoza-Araya, R.~Tonkoski, U.~Tamrakar, N.~Hatziargyriou, D.~Lagos
  \emph{et~al.}, ``Microgrid stability definitions, analysis, and examples,''
  \emph{IEEE Transactions on Power Systems}, vol.~35, no.~1, pp. 13--29, 2019.

\bibitem{parhizi2015state}
S.~Parhizi, H.~Lotfi, A.~Khodaei, and S.~Bahramirad, ``State of the art in
  research on microgrids: A review,'' \emph{IEEE Access}, vol.~3, pp. 890--925,
  2015.

\bibitem{anderson2017increasing}
K.~H. Anderson, N.~A. DiOrio, D.~S. Cutler, and R.~S. Butt, ``Increasing
  resiliency through renewable energy microgrids,'' \emph{International Journal
  of Energy Sector Management}, vol.~2, 2017.

\bibitem{hussain2019impact}
A.~Hussain, A.~O. Rousis, I.~Konstantelos, G.~Strbac, J.~Jeon, and H.-M. Kim,
  ``Impact of uncertainties on resilient operation of microgrids: A data-driven
  approach,'' \emph{IEEE Access}, vol.~7, pp. 14\,924--14\,937, 2019.

\bibitem{wu2019microgrid}
X.~Wu, Z.~Wang, T.~Ding, X.~Wang, Z.~Li, and F.~Li, ``Microgrid planning
  considering the resilience against contingencies,'' \emph{IET Generation,
  Transmission \& Distribution}, vol.~13, no.~16, pp. 3534--3548, 2019.

\bibitem{liu2017robust}
G.~Liu, M.~Starke, B.~Xiao, and K.~Tomsovic, ``Robust optimisation-based
  microgrid scheduling with islanding constraints,'' \emph{IET Generation,
  Transmission \& Distribution}, vol.~11, no.~7, pp. 1820--1828, 2017.

\bibitem{dehghani2021cyber}
M.~Dehghani, T.~Niknam, M.~Ghiasi, N.~Bayati, and M.~Savaghebi, ``Cyber-attack
  detection in dc microgrids based on deep machine learning and wavelet
  singular values approach,'' \emph{Electronics}, vol.~10, no.~16, p. 1914,
  2021.

\bibitem{wang2019deep}
H.~Wang, J.~Ruan, Z.~Ma, B.~Zhou, X.~Fu, and G.~Cao, ``Deep learning aided
  interval state prediction for improving cyber security in energy internet,''
  \emph{Energy}, vol. 174, pp. 1292--1304, 2019.

\bibitem{dehghani2021blockchain}
M.~Dehghani, M.~Ghiasi, T.~Niknam, A.~Kavousi-Fard, M.~Shasadeghi, N.~Ghadimi,
  and F.~Taghizadeh-Hesary, ``Blockchain-based securing of data exchange in a
  power transmission system considering congestion management and social
  welfare,'' \emph{Sustainability}, vol.~13, no.~1, p.~90, 2021.

\bibitem{sahoo2018stealth}
S.~Sahoo, S.~Mishra, J.~C.-H. Peng, and T.~Dragi{\v{c}}evi{\'c}, ``A stealth
  cyber-attack detection strategy for dc microgrids,'' \emph{IEEE Transactions
  on Power Electronics}, vol.~34, no.~8, pp. 8162--8174, 2018.

\bibitem{che2019impact}
L.~Che, X.~Liu, Z.~Shuai, and J.~Zhao, ``The impact of ramp-induced data
  attacks on power system operational security,'' \emph{IEEE Transactions on
  Industrial Informatics}, vol.~15, no.~9, pp. 5064--5075, 2019.

\bibitem{habibi2020detection}
M.~R. Habibi, H.~R. Baghaee, T.~Dragiˇcevi, F.~Blaabjerg \emph{et~al.},
  ``Detection of false data injection cyber-attacks in dc microgrids based on
  recurrent neural networks,'' \emph{IEEE Journal of Emerging and Selected
  Topics in Power Electronics}, 2020.

\bibitem{kushal2018risk}
T.~R.~B. Kushal, K.~Lai, and M.~S. Illindala, ``Risk-based mitigation of load
  curtailment cyber attack using intelligent agents in a shipboard power
  system,'' \emph{IEEE Transactions on Smart Grid}, vol.~10, no.~5, pp.
  4741--4750, 2018.

\bibitem{banerjee2021online}
S.~Banerjee and A.~Chatterjee, ``Online fast detection and diagnosis of power
  grid security attacks using state checksums,'' in \emph{2021 IEEE 27th
  International Symposium on On-Line Testing and Robust System Design
  (IOLTS)}.\hskip 1em plus 0.5em minus 0.4em\relax IEEE, 2021, pp. 1--7.

\bibitem{sun2018cyber}
C.-C. Sun, A.~Hahn, and C.-C. Liu, ``Cyber security of a power grid:
  State-of-the-art,'' \emph{International Journal of Electrical Power \& Energy
  Systems}, vol.~99, pp. 45--56, 2018.

\bibitem{he2021tri}
H.~He, S.~Huang, Y.~Liu, and T.~Zhang, ``A tri-level optimization model for
  power grid defense with the consideration of post-allocated dgs against
  coordinated cyber-physical attacks,'' \emph{International Journal of
  Electrical Power \& Energy Systems}, vol. 130, p. 106903, 2021.

\bibitem{gan2014convex}
L.~Gan and S.~H. Low, ``Convex relaxations and linear approximation for optimal
  power flow in multiphase radial networks,'' in \emph{2014 Power Systems
  Computation Conference}.\hskip 1em plus 0.5em minus 0.4em\relax IEEE, 2014,
  pp. 1--9.

\bibitem{nazir2020optimal}
N.~Nazir, P.~Racherla, and M.~Almassalkhi, ``Optimal multi-period dispatch of
  distributed energy resources in unbalanced distribution feeders,'' \emph{IEEE
  Transactions on Power Systems}, vol.~35, no.~4, pp. 2683--2692, 2020.

\bibitem{bolognani2015fast}
S.~Bolognani and F.~D{\"o}rfler, ``Fast power system analysis via implicit
  linearization of the power flow manifold,'' in \emph{2015 53rd Annual
  Allerton Conference on Communication, Control, and Computing
  (Allerton)}.\hskip 1em plus 0.5em minus 0.4em\relax IEEE, 2015, pp. 402--409.

\bibitem{parisio2014model}
A.~Parisio, E.~Rikos, and L.~Glielmo, ``A model predictive control approach to
  microgrid operation optimization,'' \emph{IEEE Transactions on Control
  Systems Technology}, vol.~22, no.~5, pp. 1813--1827, 2014.

\bibitem{zhang2013robust}
Y.~Zhang, N.~Gatsis, and G.~B. Giannakis, ``Robust energy management for
  microgrids with high-penetration renewables,'' \emph{IEEE transactions on
  sustainable energy}, vol.~4, no.~4, pp. 944--953, 2013.

\bibitem{bai2015robust}
X.~Bai, L.~Qu, and W.~Qiao, ``{Robust AC optimal power flow for power networks
  with wind power generation},'' \emph{IEEE Transactions on Power Systems},
  vol.~31, no.~5, pp. 4163--4164, 2015.

\bibitem{brahma2020optimal}
S.~Brahma, N.~Nazir, H.~Ossareh, and M.~Almassalkhi, ``Optimal and resilient
  coordination of virtual batteries in distribution feeders,'' \emph{IEEE
  Transactions on Power Systems}, 2020.

\bibitem{kersting2006distribution}
W.~H. Kersting, \emph{Distribution system modeling and analysis}.\hskip 1em
  plus 0.5em minus 0.4em\relax CRC press, 2006.

\bibitem{chassin2008gridlab}
D.~P. Chassin, K.~Schneider, and C.~Gerkensmeyer, ``{GridLAB-D: An open-source
  power systems modeling and simulation environment},'' in \emph{Transmission
  and distribution conference and exposition, 2008. t\&d. IEEE/PES}.\hskip 1em
  plus 0.5em minus 0.4em\relax IEEE, 2008, pp. 1--5.

\bibitem{ciraci2014fncs}
S.~Ciraci, J.~Daily, J.~Fuller, A.~Fisher, L.~Marinovici, and K.~Agarwal,
  ``{FNCS: A framework for power system and communication networks
  co-simulation},'' in \emph{Proceedings of the symposium on theory of modeling
  \& simulation-DEVS integrative}, 2014, pp. 1--8.

\end{thebibliography}


\end{document}